\documentclass[journal]{IEEEtran}
\usepackage{amsmath,amsfonts}
\usepackage{algorithm}
\usepackage{algorithmic}
\usepackage{array}
\usepackage[caption=false,font=normalsize,labelfont=sf,textfont=sf]{subfig}
\usepackage{textcomp}
\usepackage{stfloats}
\usepackage{enumerate}
\usepackage{enumitem}
\usepackage{indentfirst} 
\usepackage{amssymb}
\usepackage{amsthm}  
\usepackage{booktabs}
\usepackage{bm}
\usepackage{url}
\usepackage[compress]{cite}
\usepackage{multirow}
\usepackage{verbatim}
\usepackage{graphicx}
\usepackage{xcolor}
\def\BibTeX{{\rm B\kern-.05em{\sc i\kern-.025em b}\kern-.08em
		T\kern-.1667em\lower.7ex\hbox{E}\kern-.125emX}}
\usepackage{upgreek}
\newtheoremstyle{mylemma}
{0pt}      
{0pt}      
{\normalfont} 
{\parindent}  
{\itshape}    
{:\ }         
{0pt}         
{}            

\theoremstyle{mylemma}

\begin{document}
	\title{Ultra-High Resolution Method for Multipath Within a Co-Delay-Doppler Bin in DFT-P-OCDM}
	\author{Mingxuan Han, Weile Zhang, and Feifei Gao, \textit{Fellow, IEEE}
		\thanks{Mingxuan Han, and Weile Zhang are with the MOE Key Laboratory for
			Intelligent Networks and Network Security, School of Information and Communication Engineering, Faculty of Electronic and Information Engineering, Xi'an Jiaotong University, Xi'an 710049, China (e-mail: mxhan@stu.xjtu.edu.cn; wlzhang@mail.xjtu.edu.cn).

			Feifei Gao is with the Institute for Artificial Intelligence, Tsinghua
			University, Beijing 100084, China, also with the State Key Laboratory of
			Intelligent Technologies and Systems, Tsinghua University, Beijing 100084,
			China, and also with the Beijing National Research Center for Information
			Science and Technology, Department of Automation, Tsinghua University,
			Beijing 100084, China (e-mail: feifeigao@ieee.org).
	}}


	\maketitle

	\begin{abstract}
		Communication systems can reuse their transmitted signals for sensing without dedicated radar transmissions. For an established DFT-preprocessed orthogonal chirp division multiplexing (DFT-P-OCDM) waveform, this task becomes difficult when several physical paths in a doubly selective channel fall into the same co-delay-Doppler bin. In this case, the number of resolvable delay classes inferred from the pilot may be smaller than the number of physical paths within the co-bin. This mismatch increases the difficulty of path number determination, fractional Doppler offset estimation, and channel reconstruction. We derive a pointwise relationship between the input and output in the DFT preprocessed Fresnel (DPF) domain for doubly selective channels with multiple paths within the co-delay-Doppler bin. Based on this input and output relation, we propose the two stage ultra high resolution (TSUR) framework. The first stage uses pilots of the phase progression to estimate delay, while the second stage uses the leakage samples to estimate the Doppler and the number of paths within each delay class. Furthermore, we derive CRLBs and analyze how the pilot configuration trades sensing resolution and communication recovery. Simulation results demonstrate that TSUR resolves same delay paths within a co-delay-Doppler bin, remains robust to exist delay offsets, and achieves lower fractional Doppler estimation errors than sequential extraction and off-grid baselines.
				
	\end{abstract}


	\section{Introduction}
	\IEEEPARstart{T}{he} upcoming sixth generation (6G) wireless networks are expected to support ubiquitous connectivity and large-scale sensing applications. Existing communication waveforms and distributed receiving infrastructure can be reused for environmental sensing without introducing a dedicated radar transmitter~\cite{Rahman2020PMN, Zhu2024AFDMBistatic}. The received signals contain propagation signatures associated with static scatterers and moving targets. Accordingly, a cooperative sensing receiver must determine the number of physical propagation paths and estimate their bistatic delays, Doppler shifts, and complex gains~\cite{Ni2023CSIRatio}. These propagation parameters serve as sensing observables at the front end for subsequent localization, velocity inference, target association, and tracking. In high mobility scenarios, multipath delay and rapid Doppler variation make the channel selective in both frequency and time, thereby complicating the resolution of individual propagation paths~\cite{Bemani2023AFDM}. Chirp based multicarrier waveforms have attracted considerable interest because of their inherent robustness to Doppler effects and their suitability for ISAC~\cite{Wang2023PilotOCDM}.
	
	Orthogonal chirp division multiplexing (OCDM) and affine frequency division multiplexing (AFDM) are two representative chirp-based multicarrier waveforms. OCDM multiplexes mutually orthogonal chirps through the discrete Fresnel transform~\cite{Ouyang2016OCDM}, such that each modulated carrier extends over the entire time and frequency support of an entire symbol interval~\cite{Omar2020Spectrum}. In communication-assisted sensing, Doppler shifts are both channel impairments and observables of the propagation geometry. The deterministic chirp structure of OCDM provides a fixed observation model, making it suitable for characterizing unknown multipath environments. By comparison, AFDM extends the chirp waveform through configurable affine parameters that can be selected according to the delay and Doppler spreads. These parameters reshape the channel representation in the transform domain and improve path separability~\cite{Bao2024Tradeoff}.
	
	Across different multicarrier waveforms, existing delay-Doppler estimation methods can be grouped into on-grid and off-grid formulations. For OTFS, Reference~\cite{OTFSMishra2022} estimated on-grid delay and Doppler indices from embedded-pilot shifts and determined the path number by thresholding the detected delay-Doppler taps. For OCDM, Reference~\cite{OCDMDDdomain} exploited the sparse delay-Doppler domain representation and applied threshold detection to obtain the on-grid indices. However, Doppler shifts in high mobility channels generally do not fall exactly on the discrete grid~\cite{Kumari2023TCHTP}. The off-grid methods recover fractional delay-Doppler parameters through different signal models and inference mechanisms. Reference~\cite{Jitsumatsu2025ParallelProny} developed parallel Doppler-first and delay-first Prony estimators and combined the detected delay and Doppler taps to determine the path number. Reference~\cite{Lai2024AtomicNorm} proposed an algorithm with super resolution to estimate delay and Doppler parameters, which resolves off-grid issues through atomic norm minimization in two dimensions and optimization using ADMM, thereby enabling fractional delay and Doppler estimation. Reference~\cite{Yang2024AFDMOffGridSBL} estimated integer delays on a off-grid, while fractional Doppler offsets were treated as hyperparameters and estimated through an expectation maximization procedure for AFDM. Reference~\cite{Arous2026PCTD} estimated fractional delays and Doppler shifts for AFDM and OCDM through Dirichlet kernel correlation aided by zero padding and sequential cancellation of the strongest path. These methods improve off-grid parameter recovery, but they usually treat paths resolution through sparse supports and sequential extraction. Table~\ref{tab:comparison} summarizes representative delay-Doppler estimation methods and highlights whether the co-delay-Doppler bin path issue is explicitly addressed.
		
\begin{figure}[!t]
		\centering
		\makebox[\columnwidth][c]{\includegraphics[width=3.5in]{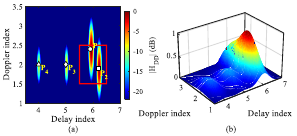}}
		\caption{DD domain channel visualization. (a) the high resolution channel magnitude and
			(b) the corresponding three dimensional representation after receiver side DD sampling. The paths $P_1$, $P_2$, $P_3$, and $P_4$ are located at the normalized continuous delay-Doppler coordinates $(d_i,\nu_i)=(5.9,2.4)$, $(6.2,1.9)$, $(5,2)$, and $(4,2)$, respectively. Although $P_1$ and $P_2$ have different fractional delay and Doppler offsets, both are mapped to the same discrete delay-Doppler bin, highlighted by the red box. }\label{figure1}
	\end{figure}

	The difficulty considered in this paper is different from off-grid parameters estimation. The wide bandwidths available at millimeter wave and terahertz frequencies provide fine delay resolution in OCDM systems~\cite{Browning2021OCDMmmWave,Li2024OCDMTHz,Wan2024OCDMmmWave}, whereas Doppler resolution remains limited by the finite coherent observation interval. In conventional ISAC systems with fixed time and frequency resources, the difference between delay and Doppler resolutions can place several physical paths in the same integer delay and Doppler bin, as illustrated in Fig.~\ref{figure1}. At the receiver, the pilot observations within a co-bin may reveal only one resolvable delay class even though multiple physical paths are present. These paths may have identical continuous delays or distinct delays whose separation is below the effective delay resolution, while having different fractional Doppler offsets. This paper refers to this situation as co-bin multipath. Resolving this case is important for practical ISAC receivers because treating one detected delay class as one physical path may cause resolvable targets within the same delay class to be missed, and degrading subsequent channel reconstruction and data recovery.
	
	Motivated by the above observations, we consider a co-bin multipath scenario and develop a TSUR framework. We first estimates the resolvable delay class. We then determines the number of physical paths represented by each resolved delay class. The main contributions are summarized as follows:
	
	\begin{table*}[t]
		\centering
		\caption{Comparison With Representative Delay-Doppler Estimation Methods}
		\label{tab:comparison}
		\renewcommand{\arraystretch}{1.10}
		\setlength{\tabcolsep}{4.2pt}
		\begin{tabular}{cccccc}
			\hline
			\textbf{Method} &
			\textbf{Signal model} &
			\textbf{Main computation} &
			\textbf{Path number} &
			\textbf{Fractional Doppler} &
			\textbf{Co-bin order mismatch} \\
			\hline
			
			\cite{OTFSMishra2022}
			& OTFS & Embedded pilot DD grid detection & thresholded DD taps & Unmodeled & Unmodeled \\
			
			\cite{OCDMDDdomain}
			& OCDM & Embedded pilot DD grid detection & thresholded DD taps & Unmodeled & Unmodeled\\
			
			\cite{Kumari2023TCHTP}
			& OTFS & TCHTP sparse recovery & thresholded DD taps & Unmodeled & Unmodeled \\

			\cite{Jitsumatsu2025ParallelProny}
			& OTFS & Parallel two stage Prony & Prony model order & Modeled & Unmodeled \\
			
			\cite{Lai2024AtomicNorm}
			& OTFS & 2D atomic norm super resolution & SIC & Modeled & Unmodeled \\
			
			\cite{Yang2024AFDMOffGridSBL}
			& AFDM & Off-grid SBL & Effective sparse support & Modeled & Unmodeled \\
			
			\cite{Arous2026PCTD}
			& AFDM/OCDM & ZP-PCTD correlation & SIC & Modeled & Unmodeled \\
			
			\textbf{This work}
			& \textbf{DFT-P-OCDM}
			& \textbf{TSUR}
			& \textbf{MDL+BIC}
			& \textbf{Modeled}
			& \textbf{Modeled} \\
			\hline
		\end{tabular}
		\par\vspace{4pt}
		
	\end{table*}

	\begin{itemize}
		\item We derive a relationship in closed form between input and output in the DPF domain for DFT-P-OCDM over doubly selective channels with joint delay and Doppler bin. The derived expression connects the channel parameters with the received pilot locations, phase progression,
		and local leakage structure. It shows that the delay is
		reflected in the pilot phase progression of the principal
		samples, whereas the fractional Doppler determines the local leakage
		pattern around the detected integer Doppler index.
		\item We propose a TSUR framework based on the derived closed form relationship between the input and output. In the first stage, TSUR extracts the principal sample from each pilot block. Because the pilot blocks are uniformly spaced, the extracted samples form an inter-pilot phase sequence governed by the continuous delays.
		The MDL criterion is applied to
		the FBSS covariance matrix to estimate the number of distinct
		delay classes, after which FBSS-MUSIC provides the initial
		delay estimates and Newton refinement improves their accuracy.
		In the second stage, TSUR extracts local leakage samples around each detected integer Doppler position. Their leakage patterns depend on the fractional Doppler offsets and are used to separate paths within the same delay class.
		For each estimated delay class, the BIC determines the number of physical paths by comparing candidate concentrated fits, and near main lobe leakage maximum likelihood estimation (NMLL-MLE) jointly estimates the corresponding fractional Doppler offsets and complex gains.
		\item We derive the CRLB for continuous delay
		and fractional Doppler estimation under the proposed observation
		model, providing theoretical benchmarks for the estimation
		accuracy. Simulations evaluate hierarchical
		path-number detection, delay and fractional Doppler estimation,
		co-bin multipath resolution, channel reconstruction, and BER under
		different SNRs, path separations, and pilot configurations. The
		results show that TSUR resolves representative co-bin multipath
		configurations, remains robust to delay mismatch within the same
		bin, and improves fractional Doppler estimation compared with
		sequential extraction and off-grid baselines. The pilot studies
		further show the sensing and communication tradeoff between co-bin
		parameter recovery, channel reconstruction, and data detection.

	\end{itemize}

	The remainder of this paper is organized as follows. Section~II introduces the DFT-P-OCDM system model and derives the DPF domain input output relationship for doubly selective channels within co delay Doppler bin. Section~III presents the ULPA frame,
	integer Doppler index estimation, and the pilot phase and local
	leakage observation models. Section~IV details the TSUR method for delay class estimation, physical path number determination, and fractional Doppler estimation. Section~V derives the CRLBs for the
	estimated channel parameters. Section~VI reports the simulation
	results on detection reliability, parameter estimation, channel
	reconstruction, BER, and pilot design tradeoff. Section~VII
	concludes the paper.

	\section{SYSTEM MODEL}
	This section introduces the DFT-P-OCDM transmitter and formulates the doubly selective channel within a co-delay-Doppler bin. It derives the DPF domain relationship between the input and output for the pilot observation models.

	\subsection{DFT-P-OCDM Transmitter}

\begin{figure*}[!t]
		\centering
		\includegraphics[width=0.90\textwidth]{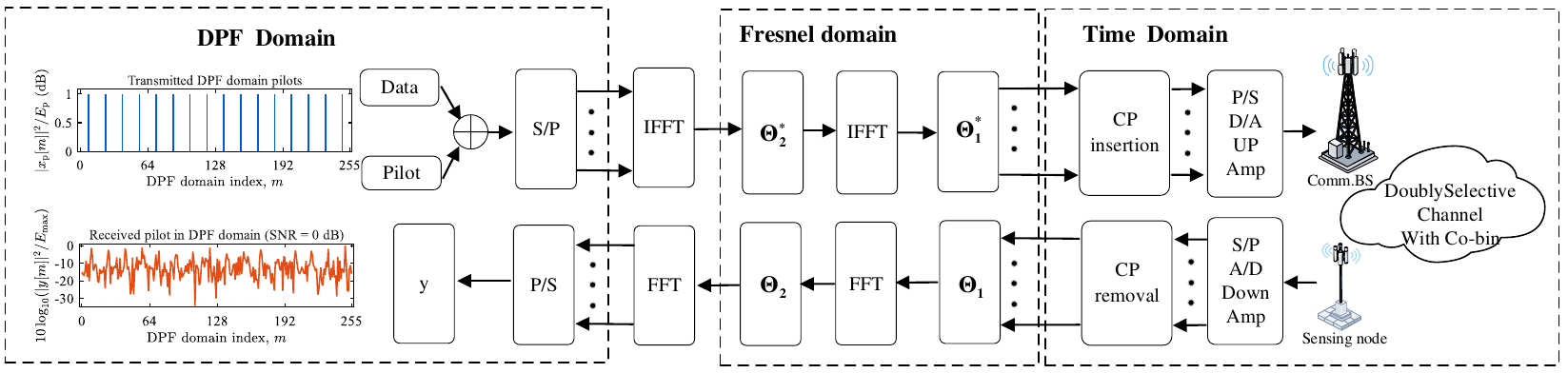}
		\caption{DFT-P-OCDM transceiver structure and ISAC node processing chain.}\label{figure2}
	\end{figure*}

	We consider an established DFT-P-OCDM transceiver with pilot and data symbols arranged in the DPF domain. The corresponding transceiver structure and processing procedure at each ISAC node are shown in Fig.~\ref{figure2}. The transmitter employs $N$ orthogonal chirp subcarriers, where $N$ is typically chosen as a power of two. 
	The input information is divided into data and pilot modules. The data information is mapped from a bit stream to independent M-ary constellation symbols. The vector of pilot and data symbols $\mathbf{x}={[x[0], x[1], \ldots, x[N-1]]}^{\mathrm{T}}$ is arranged in the DFT preprocessed Fresnel domain. It is converted into the time domain transmit vector $\mathbf{s}={[s[0], s[1], \ldots, s[N-1]]}^{\mathrm{T}}$ by successively applying the  inverse discrete Fourier transform (IDFT) and the inverse discrete Fresnel transform (IDFnT). The transmit vector in the time domain can be expressed as
	\begin{equation}
		\mathbf{s}=\boldsymbol{\Phi}^{\mathrm{H}} \mathbf{F}^{\mathrm{H}} \mathbf{x},
	\end{equation}
	where $(\cdot)^{\mathrm H}$ denotes the Hermitian transpose and
	$\mathbf{F}\in\mathbb{C}^{N\times N}$ is the normalized DFT matrix with
	$[\mathbf F]_{m,n}=N^{-1/2}e^{-\mathrm j2\pi mn/N}$ for
	$m,n=0,\ldots,N-1$.
	The matrix $\boldsymbol{\Phi}=[\boldsymbol{\varphi}_0,
	\boldsymbol{\varphi}_1,\ldots,\boldsymbol{\varphi}_{N-1}]$ denotes the
	discrete Fresnel transform (DFnT) of size $N$, where
	$\boldsymbol{\varphi}_m$ is the $m$th orthogonal chirp basis waveform. Its
	$n$th sample is
	\begin{equation}
		\varphi_m[n]=\frac{1}{\sqrt{N}}e^{-\mathrm{j}\pi/4}
		e^{\mathrm{j}\frac{\pi}{N}(m-n)^2},
		m,n=0,\ldots,N-1.
		\label{eq: dfnt_entry_v2}
	\end{equation}
	
	The mutual orthogonality of these chirp basis waveforms makes the DFnT
	matrix unitary $\boldsymbol{\Phi}^{\mathrm H}\boldsymbol{\Phi}
	=\boldsymbol{\Phi}\boldsymbol{\Phi}^{\mathrm H}=\mathbf{I}_N$. The DFnT matrix can be factorized as
	\begin{equation}
		\boldsymbol{\Phi}=\boldsymbol{\Theta}_2\mathbf{F}
		\boldsymbol{\Theta}_1,
		\label{eq:dfnt_factorization_v2}
	\end{equation}
	where $\boldsymbol{\Theta}_1=\operatorname{diag}(\boldsymbol{\theta}_1)$
	and $\boldsymbol{\Theta}_2=\operatorname{diag}(\boldsymbol{\theta}_2)$
	are diagonal chirp-phase matrices with
	$[\boldsymbol{\theta}_1]_m=e^{-\mathrm{j}\pi/4}
	e^{\mathrm{j}\pi m^2/N}$ and
	$[\boldsymbol{\theta}_2]_m=e^{\mathrm{j}\pi m^2/N}$,
	$m=0,\ldots,N-1$. According to \eqref{eq:dfnt_factorization_v2}, the IDFnT can be efficiently
	implemented using an IFFT together with two diagonal phase
	multiplications~\cite{Zhang2022OCDMCFO}. The obtained time domain symbol vector is transmitted through the antenna after D/A conversion, parallel to serial conversion, RF upconversion, and power amplification.

	\subsection{Doubly Selective Channels with Co-Delay-Doppler Bin}

	 The doubly selective channel~\cite{Guo2017HighMobility} with fractional delay and fractional Doppler is expressed as
	\begin{equation}
		\mathbf{H}
		=
		\sum_{i=0}^{L-1}
		h_i\mathbf{T}(d_i)\boldsymbol{\Delta}(\nu_i),
		\label{eq:channel_matrix}
	\end{equation}
	where $L$ is the number of propagation paths, and $h_i$, $d_i$, and $\nu_i$
	denote the complex gain, normalized delay, and normalized Doppler shift of the
	$i$th path, respectively. The fractional delay operator is then given by
	\begin{equation}
		\mathbf{T}(d_i)
		=
		\mathbf{F}^{\mathrm H}\mathbf{\Delta}(-d_i)\mathbf{F},
		\label{eq:fractional_delay_operator}
	\end{equation}
	where $\mathbf{\Delta}(-d_i)= \operatorname{diag} \left\{e^{\mathrm{-j}\frac{2\pi}{N}d_i n}\right\}_{n=0}^{N-1}$ define the diagonal phase operator. Similarly, the Doppler modulation induced by the Doppler
	shift $\nu_i$ is represented by $\boldsymbol{\Delta}(\nu_i)
	= \operatorname{diag}\left\{e^{\mathrm{j}\frac{2\pi}{N}\nu_i n}\right\}_{n=0}^{N-1}.$
	Let $\tau_i$ and $f_{D,i}$ denote the physical delay and Doppler frequency of
	the $i$th path, respectively. Their normalized forms are
	\begin{equation}
		\begin{aligned}
			d_i
			&=
			\frac{\tau_i}{T_s}
			=
			\tau_i B
			=
			l_i+\delta_i,\\
			\nu_i
			&=
			N T_s f_{D,i}
			=
			T f_{D,i}
			=
			k_i+\kappa_i,
		\end{aligned}
		\label{eq:normalized_parameters}
	\end{equation}
	where $T_s=1/B$ is the sampling interval, $B$ is the signal bandwidth,
	$T=NT_s$ is the block duration, $l_i$ and $k_i$ are the integer delay and
	Doppler indices, and $\delta_i,\kappa_i\in[-0.5,0.5)$ are the corresponding
	fractional offsets. Let
	$D_N(x)=\frac{1}{N}\sum_{n=0}^{N-1}
	e^{\mathrm{j}\frac{2\pi}{N}nx}$ denote the Dirichlet
	function. The channel response for each time domain sample can be written as
	\begin{equation}
		\begin{aligned}
			H(r,t)
			={}&
			\sum_{i=0}^{L-1}
			h_i e^{\mathrm{j}\frac{2\pi}{N}(k_i+\kappa_i)t} D_N(r-t-d_i),
		\end{aligned}
		\label{eq:time_channel_response}
	\end{equation}
	
	The fractional delay Dirichlet function has the periodic impulse representation
	\begin{equation}
		D_N(r-t-l_i-\delta_i)
		=
		\sum_{p=0}^{N-1}
		\Lambda_N(p,\delta_i)
		\delta_N[r-t-l_i-p],
		\label{eq:fractional_delay_impulse_expansion}
	\end{equation}
	and the modulation associated with the fractional Doppler admits the finite Fourier expansion $e^{\mathrm{j}\frac{2\pi}{N}\kappa_i t}=\sum_{q=0}^{N-1}
		\Gamma_N(q,\kappa_i)
		e^{\mathrm{j}\frac{2\pi}{N}qt}.
		\label{eq:fractional_doppler_fourier_expansion}$
	Where $\Lambda_N(p,\delta_i)=D_N(p-\delta_i)$,
	$\Gamma_N(q,\kappa_i)=D_N(\kappa_i-q)$, and $\delta_N[\cdot]$
	denotes the Kronecker delta with period $N$. Substituting the impulse representation of the fractional delay Dirichlet function and the finite Fourier expansion of the
	fractional Doppler modulation into \eqref{eq:time_channel_response} gives
	\begin{equation}
		\begin{aligned}
			H(r,t)
			={}&
			\sum_{i=0}^{L-1}
			\sum_{p=0}^{N-1}
			\sum_{q=0}^{N-1}
			h_i\Lambda_N(p,\delta_i)\Gamma_N(q,\kappa_i)\\
			&\times
			e^{\mathrm{j}\frac{2\pi}{N}(k_i+q)t}
			\delta_N[r-t-l_i-p].
		\end{aligned}
		\label{eq:time_channel_response_expanded}
	\end{equation}
	
	The fractional delay spreads each path over the delay samples indexed by $p$, whereas the
	fractional Doppler spreads it over the Doppler indexed by $q$. For a selected co-delay-Doppler bin with integer delay index $l_c$
	and integer Doppler index $k_c$, define the path set as
	$\mathcal C_c=\{i:l_i=l_c,\;k_i=k_c\}$. For each
	$i\in\mathcal C_c$, the continuous coordinates are
	$(d_i,\nu_i)=(l_c+\delta_i,\;k_c+\kappa_i)$. Paths in the co-bin share the same integer delay and Doppler indices. Let $d_1<d_2<\cdots<d_{K_d}$ denote the distinct continuous delays among the paths in $\mathcal C_c$. The paths with identical or insufficiently separated continuous delays may not
	be resolved individually from the pilot observation. Such paths are represented by the same
	resolved delay class. Let $d_g$ denote the representative delay of the $g$th resolved delay class,
	and let $\mathcal C_{c,g}$ denote the set of physical paths represented by this class. Let $K_g$ denote the number of physical paths in this group. The total number of physical paths in the selected co-bin is $K_c=\sum_{g=1}^{K_d}K_g$. At the receiver, the closely spaced physical delay groups may be represented by the same resolvable delay class in the pilot observation.
	
	\subsection{Input-Output Relationship of DFT-P-OCDM}
	The matrix expression of the received time domain signal is
	$\mathbf r=\mathbf H\mathbf s+\mathbf w$, where
	$\mathbf s=\boldsymbol{\Phi}^{\mathrm H}\mathbf F^{\mathrm H}\mathbf x$ and $\mathbf{w}$ is
	an AWGN with variance $\sigma^2$. The receiver applies the DFnT and then the DFT to obtain the received vector in the DPF domain as
	\begin{equation}
		\begin{aligned}
			\mathbf y
			=\mathbf F\boldsymbol{\Phi}\mathbf r
			=\mathbf F\boldsymbol{\Phi}\mathbf H
			\boldsymbol{\Phi}^{\mathrm{H}} \mathbf{F}^{\mathrm{H}}\mathbf x
			+\mathbf F\boldsymbol{\Phi} \mathbf w
			=\mathbf H_{\mathrm{DPF}}\mathbf x+\widetilde{\mathbf w},
		\end{aligned}
		\label{eq:dpf_matrix_io}
	\end{equation}
	where $\mathbf H_{\mathrm{DPF}}=\mathbf F\boldsymbol{\Phi}\mathbf H
	\boldsymbol{\Phi}^{\mathrm{H}} \mathbf{F}^{\mathrm{H}}$ and
	$\widetilde{\mathbf w}=\mathbf F\boldsymbol{\Phi}\mathbf w$.
	Since $\mathbf F$ and $\boldsymbol{\Phi}$ are unitary,
	$\widetilde{\mathbf w}$ is
	an AWGN with variance $\sigma^2$ that has the same statistical properties as $\mathbf{w}$. The expression of the channel in the Fresnel domain is $\mathbf{H}_{\mathrm{DFnT}}=\boldsymbol{\Phi} \mathbf{H} \boldsymbol{\Phi}^{\mathrm{H}}$. The channel matrix in the DPF domain and the channel matrix in the Fresnel domain satisfy the relationship $\mathbf{H}_{\mathrm{DPF}}=\mathbf{F} \mathbf{H}_{\mathrm{DFnT}} \mathbf{F}^{\mathrm{H}}$. The matrix entry of the Fresnel domain channel is given by
	\begin{equation}
		H_{\mathrm{DFnT}}(r,t)
		=
		\sum_{l=0}^{N-1}
		\sum_{k=0}^{N-1}
		\Phi(r,l)H(l,k)\Phi^{\mathrm H}(k,t).
		\label{eq:time_channel_chirp_response}
	\end{equation}
	
	Substituting \eqref{eq: dfnt_entry_v2} and
	\eqref{eq:time_channel_response_expanded} into
	\eqref{eq:time_channel_chirp_response} yields

	\begin{equation}
		\begin{aligned}
			H_{\mathrm{DFnT}}(r,t)
			={}&\frac{1}{N}
			\sum_{i=0}^{L-1}
			\sum_{p=0}^{N-1}
			\sum_{q=0}^{N-1}
			h_i\Lambda_N(p,\delta_i)\Gamma_N(q,\kappa_i)\\
			&\times
			e^{\mathrm{j}\frac{\pi}{N}\left(r^2-t^2\right)}
			\sum_{l=0}^{N-1}\sum_{k=0}^{N-1}
			e^{\mathrm{j}\frac{\pi}{N}\left(l^2-2rl\right)} \\
			&\times
			e^{-\mathrm{j}\frac{\pi}{N}
			\left[k^2-2k\left(t+k_i+q\right)\right]}
			\delta_N[l-k-l_i-p].
		\end{aligned}
		\label{eq:fractional_delay_dfnT_channel}
	\end{equation}

	The periodic impulse in \eqref{eq:fractional_delay_dfnT_channel}
	enforces $l=[k+l_i+p]_N$. Substituting this constraint into
	\eqref{eq:fractional_delay_dfnT_channel} gives
		\begin{equation}
		\begin{aligned}
			H_{\mathrm{DFnT}}(r,t)
			={}&
			\sum_{i=0}^{L-1}
			\sum_{p=0}^{N-1}
			\sum_{q=0}^{N-1}
			h_i\Lambda_N(p,\delta_i)\Gamma_N(q,\kappa_i)\\
			&\times
			e^{\mathrm{j}\frac{2\pi}{N}r(k_i+q)}
			e^{-\mathrm{j}\frac{2\pi}{N}(l_i+p)(k_i+q)}\\
			&\times
			e^{-\mathrm{j}\frac{\pi}{N}(k_i+q)^2}
			\delta_N[r-t-l_i-p-k_i-q].
		\end{aligned}
		\label{eq:dfnt_fractional_leakage_response}
	\end{equation}

	Each $(p,q)$ produces a periodic DFnT domain shift
	$l_i+p+k_i+q$, weighted jointly by the fractional delay and
	fractional Doppler leakage coefficients. The $(m,n)$ matrix entry of the DPF domain channel is obtained from the Fresnel domain channel entries as
	\begin{equation}
		H_{\mathrm{DPF}}(m,n)
		=
		\sum_{r=0}^{N-1}
		\sum_{t=0}^{N-1}
		F(m,r)H_{\mathrm{DFnT}}(r,t)
		F^{\mathrm H}(t,n).
		\label{eq:dpf_from_dfnt}
	\end{equation}
	
	Substituting normalized DFT matrix entries
	$F(m,r)=N^{-1/2}e^{-\mathrm{j}2\pi mr/N}$, conjugate transpose terms
	$F^{\mathrm H}(t,n)=N^{-1/2}
	e^{\mathrm{j}2\pi tn/N}$, and
	\eqref{eq:dfnt_fractional_leakage_response} into
	\eqref{eq:dpf_from_dfnt} gives
	\begin{equation}
		\begin{aligned}
			&H_{\mathrm{DPF}}(m,n)\\
			&=
			\sum_{i=0}^{L-1}
			\sum_{p=0}^{N-1}
			\sum_{q=0}^{N-1}
			h_i\Lambda_N(p,\delta_i)\Gamma_N(q,\kappa_i)e^{-\mathrm{j}\frac{2\pi}{N}s_{i,p}b_{i,q}}
			e^{-\mathrm{j}\frac{\pi}{N}b_{i,q}^{2}}\\
			&\times
			\frac{1}{N}
			\sum_{r=0}^{N-1}\sum_{t=0}^{N-1}e^{-\mathrm{j}\frac{2\pi}{N}mr}
			e^{\mathrm{j}\frac{2\pi}{N}r b_{i,q}}e^{\mathrm{j}\frac{2\pi}{N}tn}
			\delta_N[r-t-s_{i,p}-b_{i,q}].
		\end{aligned}
		\label{eq:dpf_fractional_leakage_response}
	\end{equation}
	where $s_{i,p}=l_i+p$ combines the integer delay index $l_i$
	with the fractional delay leakage index $p$, and
	$b_{i,q}=k_i+q$ combines the integer Doppler index $k_i$
	with the fractional Doppler leakage index $q$.
	The periodic impulse in \eqref{eq:dpf_fractional_leakage_response}
	enforces $r=[t+s_{i,p}+b_{i,q}]_N$. Applying this constraint to
	the sum over $r$ and collecting the terms that depend on $t$ give
	\begin{equation}
		\begin{aligned}
			H_{\mathrm{DPF}}(m,n)
			={}&
			\sum_{i=0}^{L-1}\sum_{p=0}^{N-1}\sum_{q=0}^{N-1}
			h_i\Lambda_N(p,\delta_i)\Gamma_N(q,\kappa_i)\\
			&\times
			e^{-\mathrm{j}\frac{2\pi}{N}m(s_{i,p}+b_{i,q})}
			e^{\mathrm{j}\frac{\pi}{N}b_{i,q}^{2}}\\
			&\times
			\frac{1}{N}\sum_{t=0}^{N-1}
			e^{-\mathrm{j}\frac{2\pi}{N}t(m-n-b_{i,q})}.
		\end{aligned}
		\label{eq:dpf_linear_sum}
	\end{equation}
	
	To evaluate the finite sum in \eqref{eq:dpf_linear_sum}, define
	$\chi_{i,q}(m,n)=\frac{1}{N}\sum_{t=0}^{N-1}
	e^{-\mathrm{j}\frac{2\pi}{N}t(m-n-b_{i,q})}$.
	Since $m$, $n$, and $b_{i,q}$ are integer-valued indices, the
	periodic orthogonality of the complex exponentials yields
	\begin{equation}
		\chi_{i,q}(m,n)
		=
		\begin{cases}
			1, & [m-n]_N=[b_{i,q}]_N,\\
			0, & \text{otherwise}.
		\end{cases}
		\label{eq:periodic_impulse_selector}
	\end{equation}
	
	Substituting \eqref{eq:periodic_impulse_selector} into
	\eqref{eq:dpf_linear_sum} gives the closed form expression for the
	$(m,n)$ entry of the channel matrix in DPF domain as
	\begin{equation}
		\begin{aligned}
			H_{\mathrm{DPF}}(m,n)
			={}&
			\sum_{i=0}^{L-1}\sum_{p=0}^{N-1}\sum_{q=0}^{N-1}
			h_i\Lambda_N(p,\delta_i)\Gamma_N(q,\kappa_i)\\
			&\times
			e^{-\mathrm{j}\frac{2\pi}{N}m(s_{i,p}+b_{i,q})}
			e^{\mathrm{j}\frac{\pi}{N}b_{i,q}^{2}}
			\delta_N[m-n-b_{i,q}].
		\end{aligned}
		\label{eq:dpf_closed_form}
	\end{equation}
	
	Equation~\eqref{eq:dpf_closed_form} separates the two leakage
	effects in the DPF domain. The $b_{i,q}$ determines the
	cyclic displacement between the input and output indices, whereas
	$s_{i,p}$ contributes to the phase rotation of the corresponding
	displaced component. Substituting \eqref{eq:dpf_closed_form} into
	\eqref{eq:dpf_matrix_io} and using the periodic impulse to evaluate
	the sum over $n$ yields the relationship between the input and output
	in the DPF domain at sample $m$ as
	\begin{equation}
		\begin{aligned}
			y_{\mathrm{DPF}}(m)
			={}&
			\sum_{i=0}^{L-1}\sum_{p=0}^{N-1}\sum_{q=0}^{N-1}
			h_i\Lambda_N(p,\delta_i)\Gamma_N(q,\kappa_i)e^{\mathrm{j}\frac{\pi}{N}b_{i,q}^{2}}\\
			&\times
			e^{-\mathrm{j}\frac{2\pi}{N}m(s_{i,p}+b_{i,q})}
			x\!\left([m-b_{i,q}]_N\right)
			+\widetilde{w}(m).
		\end{aligned}
		\label{eq:dpf_pointwise_io}
	\end{equation}
	
	For comparison, the matrix relation between the input and output of OCDM in the DFnT domain is
	$\mathbf y_{\mathrm{DFnT}}=
		\mathbf H_{\mathrm{DFnT}}\mathbf x
		+\widehat{\mathbf w},$ where $\mathbf H_{\mathrm{DFnT}}
	=\boldsymbol{\Phi}\mathbf H\boldsymbol{\Phi}^{\mathrm H}$ and
	$\widehat{\mathbf w}=\boldsymbol{\Phi}\mathbf w$. The corresponding relationship between the input and output is
	\begin{equation}
		\begin{aligned}
			y_{\mathrm{DFnT}}(r)
			={}&
			\sum_{i=0}^{L-1}\sum_{p=0}^{N-1}\sum_{q=0}^{N-1}
			h_i\Lambda_N(p,\delta_i)\Gamma_N(q,\kappa_i)\\
			&\times
			e^{-\mathrm{j}\frac{\pi}{N}
			\left(2s_{i,p}b_{i,q}+b_{i,q}^{2}\right)}
			e^{\mathrm{j}\frac{2\pi}{N}r b_{i,q}}\\
			&\times
			x\!\left([r-s_{i,p}-b_{i,q}]_N\right)
			+\widehat{w}(r),
		\end{aligned}
		\label{eq:dfnt_pointwise_io}
	\end{equation}
	where $\widehat{w}(r)$ is the $r$th sample of
	$\widehat{\mathbf w}$. Comparing \eqref{eq:dpf_pointwise_io} with \eqref{eq:dfnt_pointwise_io}
	shows a key difference in the OCDM system. Each input sample is shifted
	by $s_{i,p}+b_{i,q}$, which equals the sum of the delay related index and
	the Doppler related index. In contrast, DFT-P-OCDM imposes only the $b_{i,q}$ shift on input symbols and retains $s_{i,p}$ within the phase superposition. The DPF domain formulation translates delay into a phase offset and Doppler into a circular shift, which naturally disentangles the coupling between delay and Doppler.

	\section{ULPA FRAME AND OBSERVATION MODEL}
	This section develops the DPF domain pilot frame and the two observation models used by TSUR. The first model uses the phase variation across pilot for delay class estimation, while the second uses local fractional Doppler leakage for number of physical paths and fractional Doppler estimation within each delay class.
	\subsection{Pilot Arrangement and Integer Doppler Index Estimation}
	Let $\zeta$ protected pilot blocks be distributed uniformly over
	the DPF domain frame. The center of the $\xi$th pilot block is
	\begin{equation}
		m_{\mathrm{pilot},\xi}
		=
		m_{\mathrm{pilot},0}+\xi D_{\mathrm{pilot}},
		\xi=0,1,\ldots,\zeta-1,
		\label{eq:pilot_centers}
	\end{equation}
	where $D_{\mathrm{pilot}}$ is the interval between adjacent pilot
	centers. The pilot position and energy are known at the
	receiver. The embedded sequence of the $\xi$th pilot is
	\begin{equation}
		x_{\mathrm{pilot},\xi}[n]
		=
		\sqrt{E_{\mathrm{pilot},\xi}}\,
		\delta_N[n-m_{\mathrm{pilot},\xi}],
		\label{eq:embedded_pilot}
	\end{equation}
	where $E_{\mathrm{pilot},\xi}$ denote its energy. The corresponding pilot SNR is
	$\mathrm{SNR}_{\mathrm{pilot},\xi}
	=E_{\mathrm{pilot},\xi}/\sigma^2$. Define the protected block around the $\xi$th pilot as
	$\mathcal B_{\mathrm{pilot},\xi}
	=\{m_{\mathrm{pilot},\xi}-G,\ldots,
	m_{\mathrm{pilot},\xi}+G\}$, where $G$ is half the guard width.
	The transmitted DPF domain frame is arranged as
	\begin{equation}
		x[n]
		=
		\begin{cases}
			\sqrt{E_{\mathrm{pilot},\xi}}\,
			& n=m_{\mathrm{pilot},\xi},\\
			0,
			& n\in\mathcal B_{\mathrm{pilot},\xi}
			\setminus\{m_{\mathrm{pilot},\xi}\},\\
			x_{\mathrm d}[n],
			& n\in\mathcal D,
		\end{cases}
		\label{eq:pilot_block_arrangement}
	\end{equation}
	where $\mathcal D=\mathbb Z_N\setminus
	\bigcup_{\xi=0}^{\zeta-1}\mathcal B_{\mathrm{pilot},\xi}$ is the
	data index, where $\mathbb{Z}_N = \{0,1,\ldots,N-1\}$. Each protected block contains one known
	pilot at its center and $2G$ zero guard positions~\cite{Zhang2017ScatteredPilots}. The pilot placement must cover both positive and negative integer
	Doppler shifts and avoid overlap between adjacent guarded pilot blocks.
	The placement parameters should satisfy
	$G\ge k_{\max}$,
	$D_{\mathrm{pilot}}\ge 2G+1$,
	$m_{\mathrm{pilot},0}-G\ge0$, and
	$m_{\mathrm{pilot},0}+(\zeta-1)D_{\mathrm{pilot}}+G\le N-1$,
	where $k_{\max}=\max_i |k_i|$. The number of data positions is
	$N_{\mathrm d}=N-\zeta(2G+1)$, and the total transmitted pilot
	energy is
	$E_{\mathrm{pilot}}
	=\sum_{\xi=0}^{\zeta-1}E_{\mathrm{pilot},\xi}$.
	Without prior information on the relative importance of different pilot blocks, the pilot energy is uniformly allocated as
	$E_{\mathrm{pilot},\xi}=E_{\mathrm{pilot}}/\zeta$.
	Consequently, $\zeta$, $G$, and $E_{\mathrm{pilot}}$ determine the resource tradeoff between sensing and communication.

	\textbf{Estimation of the integer Doppler index:} The transmitted pilot centers and energies defined in
	\eqref{eq:pilot_centers}-\eqref{eq:pilot_block_arrangement} are
	known at the receiver. According to
	\eqref{eq:dpf_pointwise_io}, the main component associated with
	the $i$th path is displaced from a transmitted pilot center by
	the integer Doppler $k_i$. The receiver performs cross correlation with the known transmitted pilot using a threshold~\cite{Mahfoudia2020PilotPCL}. The displacement of each retained correlation peak
	from $m_{\mathrm{pilot},0}$ provides the corresponding integer
	Doppler estimate. 
	
	\subsection{Pilot Phase Model for Delay}
	The fractional delay satisfies $\sum_{p=0}^{N-1}\Lambda_N(p,\delta_i)
		e^{-\mathrm{j}2\pi mp/N}
		=
		e^{-\mathrm{j}2\pi m\delta_i/N}.$
	Applying this identity eliminates the leakage index $p$. The relationship between the input and output in the DPF domain is
	\begin{equation}
		\begin{aligned}
			y_{\mathrm{DPF}}(m)
			={}&
			\sum_{i=0}^{L-1}\sum_{q=0}^{N-1}
			h_i\Gamma_N(q,\kappa_i)
			e^{\mathrm{j}\frac{\pi}{N}(k_i+q)^2}\\
			&\times
			e^{-\mathrm{j}\frac{2\pi}{N}m(d_i+k_i+q)}
			x\!\left([m-k_i-q]_N\right)
			+\widetilde{w}(m).
		\end{aligned}
		\label{eq:dpf_delay_collapsed_io}
	\end{equation}

	Differences in their fractional Doppler offsets produce distinct leakage patterns with period $N$ around the common displacement. For the $\xi$th received pilot block, let
	$z\in\mathbb Z_N$ denote the sample offset from its response
	center. The corresponding DPF output index is $m_{\xi,z}=[m_{\mathrm{pilot},\xi}+k_c+z]_N.$
	For the $i$th path and the $\upsilon$th transmitted pilot, define
	\begin{equation}
		q_{i,\xi,\upsilon}(z)
		=
		[z+(\xi-\upsilon)D_{\mathrm{pilot}}+k_c-k_i]_N,
		\upsilon=0,1,\ldots,\zeta-1,
		\label{eq:pilot_leakage_index}
	\end{equation}
	where $q_{i,\xi,\upsilon}(z) = q $ is the periodic leakage index
	associated with the $i$th path. The additional index
	$\upsilon$ denotes the transmitted pilot block whose contribution is
	observed in the $\xi$th received pilot window. It follows from
	\eqref{eq:pilot_centers} that
	$[m_{\xi,z}-k_i-q_{i,\xi,\upsilon}(z)]_N
	=m_{\mathrm{pilot},\upsilon}$. The $\upsilon=\xi$ corresponds to
	the current pilot contribution, whereas $\upsilon\ne\xi$ represents leakage between different pilots from other transmitted pilot blocks. For the current pilot contribution with $\upsilon=\xi$ and $k_i=k_c$,
	$q_{i,\xi,\xi}(z)=[z]_N$. The local coordinate $z$ directly
	indexes the fractional Doppler leakage around the principal response. The local observation in the $\xi$th received pilot block is denoted by
	$y_\xi(z)$.
	Substituting the protected pilot sequence from
	\eqref{eq:pilot_leakage_index} into
	\eqref{eq:dpf_delay_collapsed_io} gives
	\begin{equation}
		\begin{aligned}
			y_\xi(z)
			={}&
			\sum_{i=0}^{L-1}
			\sum_{\upsilon=0}^{\zeta-1}
			h_i\sqrt{E_{\mathrm{pilot},\upsilon}}\,
			\Gamma_N\!\left(q_{i,\xi,\upsilon}(z),\kappa_i\right)e^{-\mathrm{j}\frac{2\pi}{N}m_{\xi,z}d_i}\\
			&\times
			e^{-\mathrm{j}\frac{\pi}{N} [2m_{\xi,z}-k_i-q_{i,\xi,\upsilon}(z)]
				[k_i+q_{i,\xi,\upsilon}(z)]}
			+\widetilde{w}(z).
		\end{aligned}
		\label{eq:exact_multi_pilot_observation}
	\end{equation}
	
	Equation~\eqref{eq:exact_multi_pilot_observation} includes the
	pilot induced contributions of all $L$ paths and all $\zeta$
	transmitted pilots. A path with $k_i\ne k_c$ is not discarded.
	Its response enters through the offset $k_c-k_i$ in
	$q_{i,\xi,\upsilon}(z)$. At the principal sample
	$z=0$, the desired current pilot term satisfies $\upsilon=\xi$ and
	$q_{i,\xi,\xi}(0)=0$ for every $i\in\mathcal C_c$. Within the
	integer Doppler group $k_i=k_c$, the contribution from another
	pilot block is weighted by
	$\Gamma_N([(\xi-\upsilon)D_{\mathrm{pilot}}]_N,\kappa_i)$.
	The pilot is selected to satisfy
	\begin{equation}
		\max_{\xi,\,\kappa}
			\frac{
				\displaystyle\sum_{\upsilon\ne\xi}
				E_{\mathrm{pilot},\upsilon}
				\left|\Gamma_N\!\left(
				[(\xi-\upsilon)D_{\mathrm{pilot}}]_N,\kappa
				\right)\right|^2
			}{
				E_{\mathrm{pilot},\xi}
				\left|\Gamma_N(0,\kappa)\right|^2
			}\\[-0.25ex]
			\le \varepsilon_{\mathrm{pilot}}.
		\label{eq:cross_pilot_leakage_condition}
	\end{equation}
	
	Where $\varepsilon_{\mathrm{pilot}}\ll1$ is the prescribed upper bound
	on the leakage between different pilots. Since $k_c$ is known after integer Doppler index, the only
	unknown parameter in this pilot phase progression is the
	continuous normalized delay $d_i=l_i+\delta_i$. This structure forms
	the delay observation model used by the subsequent TSUR estimator. Following \eqref{eq:cross_pilot_leakage_condition}, the reduced delay
	observation keeps only the principal term of the current pilot with
	$\upsilon=\xi$ and $z=0$. The remaining terms are treated as residual
	components. For every $i\in\mathcal C_c$, the phase affected by delay
	in \eqref{eq:exact_multi_pilot_observation} is
	$e^{-\mathrm{j}2\pi m_{\xi,0}d_i/N}$. The phase can therefore be written as
	$e^{-\mathrm{j}2\pi m_{\mathrm{pilot},0}(d_i+k_c)/N}
	[e^{-\mathrm{j}2\pi D_{\mathrm{pilot}}(d_i+k_c)/N}]^\xi$.
	Since $k_c$ is known after integer Doppler index, the only
	unknown parameter in this pilot phase progression is the
	continuous normalized delay $d_i=l_i+\delta_i$. This structure forms
	the delay observation model used by the subsequent TSUR estimator.
	
	\subsection{Local Leakage Model for Fractional Doppler}
	Fractional Doppler offsets are estimated from the leakage around the detected integer Doppler index. The local observation window is centered at the detected integer Doppler
	index, with the same number of retained samples on its left and right. Based on \eqref{eq:exact_multi_pilot_observation}, the samples within this local window define a reduced observation that contains the desired co-bin components and the dominant structured leakage entering the window. Using the pilot centers in \eqref{eq:pilot_centers} and the leakage
	index in \eqref{eq:pilot_leakage_index}, define the pilot component
	response of $i$th path induced by the $\upsilon$th transmitted pilot as
	\begin{equation}
		\begin{aligned}
			b_{i,\xi,\upsilon}(z)
			={}&
			\sqrt{E_{\mathrm{pilot},\upsilon}}\,
			\Gamma_N\!\left(q_{i,\xi,\upsilon}(z),\kappa_i\right)\\
			&\times
			e^{\mathrm{j}\frac{\pi}{N}
				[k_i+q_{i,\xi,\upsilon}(z)]^2}
			e^{-\mathrm{j}\frac{2\pi}{N}m_{\xi,z}
				[d_i+k_i+q_{i,\xi,\upsilon}(z)]}.
		\end{aligned}
		\label{eq:local_pilot_component}
	\end{equation}
	
	Substituting \eqref{eq:local_pilot_component} into \eqref{eq:exact_multi_pilot_observation} and simplifying, we obtain 
	\begin{equation}
		y_\xi(z)
		=
		\sum_{i=0}^{L-1}h_i
		\sum_{\upsilon=0}^{\zeta-1}b_{i,\xi,\upsilon}(z)
		+\widetilde{w}(z).
		\label{eq:exact_local_leakage_model}
	\end{equation}
	
	Equation~\eqref{eq:exact_local_leakage_model} is the reference local
	model before any weak structured component is absorbed into the
	residual. We use a symmetric window with $Q$
	samples on each side of $z=0$. The index set of the local window is
	$\mathcal Z_Q=\{-Q,\ldots,Q\}$, and the window contains
	$N_Q=2Q+1$ samples. The windowed observation is defined as
	$\mathbf y_\xi^{(Q)}=[y_\xi(z)]_{z\in\mathcal Z_Q}$.
	The corresponding component and noise vectors are
	$\mathbf b_{i,\xi,\upsilon}^{(Q)}
	=[b_{i,\xi,\upsilon}(z)]_{z\in\mathcal Z_Q}$ and
	$\widetilde{\mathbf w}_\xi^{(Q)}
	=[\widetilde w_\xi(z)]_{z\in\mathcal Z_Q}$, respectively. The fraction of the current pilot leakage energy retained by this window is $\eta_Q(\kappa)=\sum_{z=-Q}^{Q}
		\left|\Gamma_N([z]_N,\kappa)\right|^2, 0<\eta_Q(\kappa)\le1.$ The energy captured by a fixed $Q$ depends on the fractional Doppler offset
	$\kappa$. A prescribed retention level $\eta_0\in(0,1]$ can be
	enforced through
	\begin{equation}
		\min_{\kappa\in[-1/2,\,1/2)}
		\eta_Q(\kappa)
		\ge
		\eta_0.
		\label{eq:minimum_retained_energy}
	\end{equation}

	The same window also determines how much leakage from other pilot
	blocks enters the local observation. For the component associated with path $i$ and transmitted pilot $\upsilon$, define its energy within the window as
	\begin{equation}
		\begin{aligned}
			P_{i,\upsilon}^{(Q)}
			={}&
			\left|h_i\right|^2 E_{\mathrm{pilot},\upsilon}
			\sum_{z=-Q}^{Q}
			\left|\Gamma_N\!\left(
			q_{i,\xi,\upsilon}(z),\kappa_i
			\right)\right|^2 .
		\end{aligned}
		\label{eq:in_window_leakage_power}
	\end{equation}
	
	For the current pilot of a path in the detected integer Doppler
	group, $P_{i,\xi}^{(Q)}
	=\left|h_i\right|^2
	E_{\mathrm{pilot},\xi}\eta_Q(\kappa_i)$.
	The normalized leakage interaction metric is
	\begin{equation}
		\rho_{i,\upsilon}^{(Q)}
		=
		\frac{P_{i,\upsilon}^{(Q)}}
		{\left|h_i\right|^2
		E_{\mathrm{pilot},\xi}\eta_Q(\kappa_i)}.
		\label{eq:normalized_interaction_level}
	\end{equation}
	
	This ratio depends on $Q$, $\kappa_i$, $D_{\mathrm{pilot}}$, and the integer Doppler separation $k_c-k_i$. It characterizes the relative strength of the leakage contribution from the $\upsilon$th transmitted pilot within the local observation window of the $\xi$th received pilot. A larger value indicates a stronger contribution to the retained local observation. The \eqref{eq:normalized_interaction_level} is used to analyze the relative importance of leakage between different pilots and to motivate the reduced neighboring pilot model.
	
	All paths in the selected co-bin are retained as target components. To control the complexity of modeling leakage between different pilots, a prescribed pilot interaction radius $R$ is used. For each received pilot window, the local response model includes the contributions from the current transmitted pilot and up to $R$ neighboring pilots on each side. The remaining weaker interactions are absorbed into the residual. Suppose that the first step provides $\widehat K_d$ delay estimates
	$\widehat{\mathbf d}
	=[\widehat d_1,\ldots,\widehat d_{\widehat K_d}]^{\mathrm T}$.
	The observation in the second stage is formed from a prescribed set of received pilot windows
	$\mathcal S=\{\xi_1,\ldots,\xi_{N_{\mathrm w}}\}
	\subseteq\{0,\ldots,\zeta-1\}$.
	The set $\mathcal S$ allows flexible pilot window selection for joint
	local leakage fitting, ranging from all received pilot windows to the
	windows retained by a reduced NMLL model. Define
	\begin{equation}
		\begin{aligned}
			\overline{\mathbf y}_{\mathcal S}^{(Q)}
			=\operatorname{col}\{
			\mathbf y_{\xi}^{(Q)}:\xi\in\mathcal S\},
			N_{\mathrm{st}}=N_{\mathrm w}N_Q ,
		\end{aligned}
		\label{eq:stacked_local_observation_definition}
	\end{equation}
	where $\operatorname{col}\{\cdot\}$ denotes vertical
	concatenation. For $g$th delay class, let $K$ be a candidate physical path number
	and define
	$\boldsymbol{\kappa}_g
	=[\kappa_{g,1},\ldots,\kappa_{g,K}]^{\mathrm T}$.
	In the reduced second stage model, the paths assigned to $g$th class
	are represented by the refined delay $\widehat d_g$ and are
	distinguished by their fractional Doppler leakage responses. This representation is exact when the physical paths assigned to the same resolved delay class have an identical continuous delay. When several distinct continuous delays are too closely spaced to be resolved separately, $\hat d_g$ serves as a representative delay for the corresponding paths. Let
	$\mathbf b_\xi^{(Q)}(\widehat d_g,k_c,\kappa)$ denote the sum of
	the retained pilot component vectors associated with one
	candidate path, and define its stacked response as
	\begin{equation}
		\overline{\mathbf b}_{\mathcal S}^{(Q)}
		(\widehat d_g,k_c,\kappa)
		=
		\operatorname{col}\{
		\mathbf b_\xi^{(Q)}
		(\widehat d_g,k_c,\kappa):\xi\in\mathcal S\}.
		\label{eq:stacked_path_response}
	\end{equation}
	
	The class response dictionary associated with $g$th delay class and
	candidate order $K$ is given by
	\begin{equation}
		\begin{aligned}
			\overline{\mathbf B}_{\mathcal S,g,K}^{(Q)}
			(\boldsymbol{\kappa}_g)
			=
			\big[
			\overline{\mathbf b}_{\mathcal S}^{(Q)}
			(\widehat d_g,k_c,\kappa_{g,1}),\ldots,
			\overline{\mathbf b}_{\mathcal S}^{(Q)}
			(\widehat d_g,k_c,\kappa_{g,K})
			\big].
		\end{aligned}
		\label{eq:target_leakage_dictionary}
	\end{equation}
	
	The local observation used for number of paths
	and fractional Doppler estimation is
	\begin{equation}
		\begin{aligned}
			\overline{\mathbf y}_{\mathcal S}^{(Q)}
			={}
			\overline{\mathbf B}_{\mathcal S,\mathrm{tar}}^{(Q)}
			(\boldsymbol{\kappa})\mathbf h
			+
			\overline{\mathbf B}_{\mathcal S,\mathrm{nui}}^{(Q)}
			\mathbf h_{\mathrm{nui}}
			+\overline{\mathbf e}_{\mathcal S}^{(Q)}
			+\mathbf w_{\mathcal S}^{(Q)},
		\end{aligned}
		\label{eq:second_stage_local_observation}
	\end{equation}
	where $\overline{\mathbf B}_{\mathcal S,\mathrm{tar}}^{(Q)}$
	contains the modeled local responses of the co-bin paths
	formed with the delay estimates obtained, and
	$\mathbf h$ collects the corresponding complex gains. These responses include the principal contribution and the retained leakage structure within the window of the candidate paths. The matrix
	$\overline{\mathbf B}_{\mathcal S,\mathrm{nui}}^{(Q)}$ contains the
	retained strong structured leakage terms that are not included in
	$\overline{\mathbf B}_{\mathcal S,\mathrm{tar}}^{(Q)}$, such as
	the leakage between different pilots and strong out of bin path leakage. Their
	unknown gains are collected in $\mathbf h_{\mathrm{nui}}$. The vector $\overline{\mathbf e}_{\mathcal S}^{(Q)}$ collects the weak structured terms omitted from the reduced model. These terms include weak leakage between different pilots, weak leakage from paths outside the considered bin, and residual leakage from data symbols entering the selected pilot windows. The effect of data symbol leakage
	is generally more pronounced for received pilot windows located closer to the data region.

	\section{TSUR Framework}
	
	The TSUR framework is applied to each detected co-delay-Doppler bin and
	is developed from the two pilot observation models established in the
	preceding section. The first stage uses the principal pilot samples to
	estimate the delay classes. The second stage uses the local
	leakage samples to estimate Doppler and the number of paths.
	
	\subsection{The First Stage of the TSUR Framework}
	The first step estimates the delays from the
	phase of the pilot derived in the preceding section. Taking
	$z=0$ in~\eqref{eq:exact_multi_pilot_observation}, retaining the current pilot window $\upsilon=\xi$, and using $q_{i,\xi,\xi}(0)=0$ for the detected integer Doppler index in co-bin give
	\begin{equation}
		y_\xi(0)
		=
		\sum_{g=1}^{K_d}
		\sum_{r=1}^{K_g}
		\beta_c
		e^{-\mathrm{j}\frac{2\pi}{N}
			\xi D_{\mathrm{pilot}}(d_c+k_c)}
		+\widetilde{w},
	\end{equation}
	where $r$ denotes the index of a physical path within a delay class. Stacking
	$y_\xi(0)$ over $\xi=0,\ldots,\zeta-1$ gives
	\begin{equation}
		\mathbf y_p
		=
		\mathbf A_d(\mathbf d)\boldsymbol{\beta}
		+\widetilde{\mathbf w},
		\label{eq:first_stage_delay_observation}
	\end{equation}
	where
	$\mathbf A_d(\mathbf d)
	=[\mathbf a_d(d_1),\ldots,\mathbf a_d(d_{K_d})]$,
	$\mathbf d=[d_1,\ldots,d_{K_d}]^{\mathrm T}$, and
	$\boldsymbol{\beta}=[\beta_1,\ldots,\beta_{K_d}]^{\mathrm T}$.
	The delay steering vector is $\mathbf a_d(d)=\bigl[	1,\,e^{-\mathrm{j}\frac{2\pi D_{\mathrm{pilot}}}{N}(d+k_c)},\ldots,
		e^{-\mathrm{j}(\zeta-1)\frac{2\pi D_{\mathrm{pilot}}}{N}(d+k_c)}
		\bigr]^{\mathrm T}.$
	The response dictionary associated with the $g$ delay class and candidate path number $K$ is given by
	\begin{equation}
		\begin{aligned}
			\beta_c
			={}
			\sqrt{\frac{E_{\mathrm{pilot}}}{\zeta}}\,
			e^{\mathrm{j}\frac{\pi}{N}k_c^2}
			e^{-\mathrm{j}\frac{2\pi}{N}
				(m_{pi.,0}+k_c)(d_c+k_c)}
			\sum_{i\in\mathcal C_{c}}
			h_i\Gamma_N(0,\kappa_i).
		\end{aligned}
	\end{equation}

	Physical paths sharing the same continuous delay $d_r$ contribute to the same steering component in (37), with their complex gains and principal fractional Doppler coefficients combined in $\beta_r$. Due to finite SNR and limited pilot observations, closely spaced delays may be unresolved in the first step and represented by a single resolved delay class. 

	\textbf{1) FBSS-MUSIC Initialization:}
	The components in \eqref{eq:first_stage_delay_observation} may be
	coherent. Therefore, the forward and backward spatial smoothing is applied
	before the MUSIC decomposition. Let $M_s$ denote the subarray length used
	for smoothing, and let $J_s=\zeta-M_s+1$ denote the number of overlapping
	forward subarrays. If $\mathbf y_j$ denotes the $j$th subarray of
	$\mathbf y_p$ with length $M_s$, the smoothed covariance matrix is
	\begin{equation}
		\widehat{\mathbf R}_{\mathrm{FB}}
		=
		\frac{1}{2}
		\left(
		\widehat{\mathbf R}_{\mathrm f}
		+\mathbf J_{M_s}
		\widehat{\mathbf R}_{\mathrm f}^{*}
		\mathbf J_{M_s}
		\right),
		\label{eq:fbss_covariance}
	\end{equation}
	where
	$\widehat{\mathbf R}_{\mathrm f}
	=J_s^{-1}\sum_{j=0}^{J_s-1}\mathbf y_j\mathbf y_j^{\mathrm H}$
	is the forward-smoothed covariance matrix and
	$\mathbf J_{M_s}$ is the $M_s\times M_s$ exchange matrix.
	Let $\lambda_1\geq\cdots\geq\lambda_{M_s}$ denote the eigenvalues
	of $\widehat{\mathbf R}_{\mathrm{FB}}$. For each candidate order
	$k$, the MDL~\cite{Wax1985ITC} criterion and the selected order are
	\begin{equation}
		\begin{aligned}
			\operatorname{MDL}_d(k)
			&=
			-J_s(M_s-k)
			\log\!\left(
			\frac{
				\left(
				\prod_{m=k+1}^{M_s}\lambda_m
				\right)^{\frac{1}{M_s-k}}
			}{
				\frac{1}{M_s-k}
				\sum_{m=k+1}^{M_s}\lambda_m
			}
			\right)\\
			&\quad+
			\frac{k(2M_s-k)}{2}\log J_s,\\
			\widehat K_d
			&=
			\underset{0\leq k\leq K_{d,\max}}{\arg\min}\,
			\operatorname{MDL}_d(k).
		\end{aligned}
		\label{eq:delay_mdl_order}
	\end{equation}
	where $K_{d,\max}<M_s$ is the prescribed upper bound.
	The estimate $\widehat K_d$ denotes the number of resolvable delay
	classes, rather than the physical path number. Several physical paths
	may share one delay class. To form the noise subspace and recover coherent components, the
	smoothing parameters must satisfy
	$M_s>K_d$ and $J_s\geq K_d$.  Let $\mathbf U_n$ contain the eigenvectors
	associated with the $M_s-\widehat K_d$ smallest eigenvalues of
	$\widehat{\mathbf R}_{\mathrm{FB}}$. The FBSS-MUSIC spectrum is
	\begin{equation}
		P_{\mathrm{FBSS}}(d)
		=
		\frac{1}{
			\mathbf a_s^{\mathrm H}(d)
			\mathbf U_n\mathbf U_n^{\mathrm H}
			\mathbf a_s(d)},
		\label{eq:fbss_music_spectrum}
	\end{equation}
	where $\mathbf a_s(d)$ contains the first $M_s$ entries of
	$\mathbf a_d(d)$. The $\widehat K_d$ dominant
	peaks form the coarse delay vector
	$\widetilde{\mathbf d}
	=[\widetilde d_1,\ldots,\widetilde d_{\widehat K_d}]^{\mathrm T}$. These coarse estimates provide the initial values for the subsequent likelihood based delay refinement.

	\textbf{2) Likelihood-Based Delay Refinement:}
	The FBSS-MUSIC estimates initialize the following concentrated
	likelihood criterion~\cite{Zhang2016BlindCFO}:
	\begin{equation}
		\Omega(\mathbf d)
		=
		\operatorname{tr}\!\left[
		\mathbf P_d(\mathbf d)\widehat{\mathbf R}_{\mathrm{FB}}
		\right],
		\label{eq:delay_likelihood_criterion}
	\end{equation}
	where $\mathbf P_d(\mathbf d)
	=\mathbf A_s(\mathbf d)\mathbf A_s^\dagger(\mathbf d)$,
	$\mathbf A_s(\mathbf d)
	=[\mathbf a_s(d_1),\ldots,\mathbf a_s(d_{\widehat K_d})]$ and
	$\mathbf A_s^\dagger
	=(\mathbf A_s^{\mathrm H}\mathbf A_s)^{-1}
	\mathbf A_s^{\mathrm H}$. The criterion measures the covariance
	energy explained by the candidate delay subspace. A joint maximization of
	\eqref{eq:delay_likelihood_criterion} provides a
	maximum-likelihood benchmark when a multidimensional search is
	feasible. Alternating optimization offers a lower complexity numerical alternative. It updates one delay at a time while fixing
	the remaining delays at their latest values. In the proposed
	implementation, the FBSS-MUSIC estimates are refined jointly by
	Newton's method. The Newton update uses the local gradient and
	Hessian of the criterion to jointly refine all delay parameters without
	an exhaustive multidimensional search.

	\textbf{3) Newton refinement:}
	The refinement is initialized with
	$\mathbf d^{(0)}=\widetilde{\mathbf d}$. At iteration $t$, the stationarity condition
	$\nabla_{\mathbf d}\Omega(\widehat{\mathbf d})=\mathbf 0$ is linearized around
	$\mathbf d^{(t)}$ using a first order Taylor expansion
	\begin{equation}
		\mathbf 0
		\approx
		\nabla_{\mathbf d}\Omega(\mathbf d^{(t)})
		+
		\nabla_{\mathbf d}^{2}\Omega(\mathbf d^{(t)})
		\left(\mathbf d^{(t+1)}-\mathbf d^{(t)}\right).
	\end{equation}
	
	For $g,r\in\{1,\ldots,\widehat K_d\}$, the $g$th gradient entry and
	the
	$(g,r)$th Hessian entry are
	\begin{equation}
		\frac{\partial\Omega(\mathbf d)}{\partial d_g}
		=
		\operatorname{tr}\!\left[
		\frac{\partial\mathbf P_d(\mathbf d)}{\partial d_g}
		\widehat{\mathbf R}_{\mathrm{FB}}
		\right],
		\frac{\partial^2\Omega(\mathbf d)}
		{\partial d_g\partial d_r}
		=
		\operatorname{tr}\!\left[
		\frac{\partial^2\mathbf P_d(\mathbf d)}
		{\partial d_g\partial d_r}
		\widehat{\mathbf R}_{\mathrm{FB}}
		\right].
		\label{eq:delay_gradient_hessian}
	\end{equation}
	
	The first derivative of the projection matrix is
	\begin{equation}
		\frac{\partial\mathbf P_d}{\partial d_g}
		=
		\frac{\partial\mathbf A_s}{\partial d_g}
		\mathbf A_s^\dagger
		+
		\mathbf A_s
		\frac{\partial\mathbf A_s^\dagger}{\partial d_g}.
		\label{eq:delay_projection_first_derivative}
	\end{equation}
	
	The mixed second derivative can be expressed as
	\begin{equation}
		\frac{\partial^2\mathbf P_d}
		{\partial d_g\partial d_r}
		=
		\boldsymbol{\Xi}_{g,r}
		+\boldsymbol{\Xi}_{g,r}^{\mathrm H},
		\label{eq:delay_projection_second_derivative}
	\end{equation}
	where
	\begin{equation}
		\begin{aligned}
			\boldsymbol{\Xi}_{g,r}
			={}&
			-\mathbf P_d
			\frac{\partial\mathbf A_s}{\partial d_r}
			\mathbf A_s^\dagger
			\frac{\partial\mathbf A_s}{\partial d_g}
			\mathbf A_s^\dagger
			+
			(\mathbf A_s^\dagger)^{\mathrm H}
			\frac{\partial\mathbf A_s^{\mathrm H}}{\partial d_r}
			\mathbf P_d
			\frac{\partial\mathbf A_s}{\partial d_g}
			\mathbf A_s^{\mathrm H}\\
			&-
			\mathbf P_d
			\frac{\partial^2\mathbf A_s}
			{\partial d_g\partial d_r}
			\mathbf A_s^\dagger
			+
			\mathbf P_d
			\frac{\partial\mathbf A_s}{\partial d_g}
			(\mathbf A_s^{\mathrm H}\mathbf A_s)^{-1}
			\frac{\partial\mathbf A_s^{\mathrm H}}{\partial d_r}
			\mathbf P_d\\
			&+
			\mathbf P_d
			\frac{\partial\mathbf A_s}{\partial d_g}
			\mathbf A_s^\dagger
			\frac{\partial\mathbf A_s}{\partial d_r}
			\mathbf A_s^\dagger.
		\end{aligned}
		\label{eq:delay_projection_xi}
	\end{equation}
	
	All quantities in \eqref{eq:delay_gradient_hessian} and
	\eqref{eq:delay_projection_second_derivative} are evaluated at the
	current delay vector. The Newton update follows as
	\begin{equation}
		\mathbf d^{(t+1)}
		=
		\mathbf d^{(t)}
		-
		\left[
		\nabla_{\mathbf d}^{2}\Omega(\mathbf d^{(t)})
		\right]^{-1}
		\nabla_{\mathbf d}\Omega(\mathbf d^{(t)}).
		\label{eq:delay_newton_update}
	\end{equation}
	
	When the full step decreases
	\eqref{eq:delay_likelihood_criterion}, the step length is reduced
	by backtracking. The iteration stops when
	\begin{equation}
		\left\|
		\mathbf d^{(t+1)}-\mathbf d^{(t)}
		\right\|_2
		\leq\epsilon_d
		\quad\text{or}\quad
		t=T_{\max}.
		\label{eq:delay_newton_stopping}
	\end{equation}
	
	Both the converged Newton estimate and the FBSS-MUSIC
	initialization are evaluated using $\Omega(\mathbf d)$. If the
	initialization yields a larger objective value, it is retained as
	the final delay estimate. The retained vector is denoted by
	$\widehat{\mathbf d}
	=[\widehat d_1,\ldots,\widehat d_{\widehat K_d}]^{\mathrm T}$.

	\subsection{The Second Stage of the TSUR Framework}

	The second step determines the physical paths by each delay class and estimates the corresponding fractional Doppler offsets. For class $g$, the refined delay
	$\widehat d_g$ and the detected integer Doppler index $k_c$ are
	treated as known. The stacked local observation
	$\overline{\mathbf y}_{\mathcal S}^{(Q)}$ is constructed for fractional Doppler estimation.

	\textbf{1) selection of near principal values:}
	For every $\xi\in\mathcal S$, the principal sample associated with $k_c$ is indexed by $z=0$, and the estimator retains the symmetric window
	$\mathcal Z_Q=\{-Q,\ldots,Q\}$.
	The retained window contains the principal contribution and its dominant fractional Doppler leakage around the principal sample. Leakage between different pilot blocks is incorporated according to the prescribed pilot interaction radius $R$ defined in Section~III. For each received pilot window, the local response model includes the contribution from the current transmitted pilot together with those from up to $R$ neighboring transmitted pilots on each side. This fixed modeling extent retains the dominant pilot interactions while avoiding the need to model the complete $N$ point response.


	\textbf{2) Path Number and Fractional Doppler Estimation:}
	Let $\mathcal M=(K_1,\ldots,K_{\widehat K_d})$ denote a candidate path number model, where $K_g\geq1$ is the assumed number of physical paths associated with the $g$th delay class. The total number of paths under model $\mathcal M$ is $K_{\mathcal M}=\sum_{g=1}^{\widehat K_d}K_g$, subject to $K_{\mathcal M} \leq K_{c,\max}$.The admissible model set is denoted by $\mathfrak M$. The candidate fractional Doppler offsets in the $g$th resolved delay
	class are collected in $\boldsymbol{\kappa}_g=[\kappa_{g,1},\ldots,\kappa_{g,K_g}]^{\mathrm T},
	-\frac{1}{2}\leq \kappa_{g,1}<\cdots<\kappa_{g,K_g}<\frac{1}{2}.$ The ordering constraint removes permutation ambiguity only among paths
	represented by the same delay class. Fractional Doppler offsets belonging to
	different delay classes are allowed to be equal. A candidate model $\mathcal M$ collects the fractional Doppler offsets in
	$\boldsymbol{\kappa}
	=[\boldsymbol{\kappa}_1^{\mathrm T},\ldots,
	\boldsymbol{\kappa}_{\widehat K_d}^{\mathrm T}]^{\mathrm T}$.
	The augmented model matrix is
	\begin{equation}
		\begin{aligned}
			\overline{\mathbf A}_{\mathcal S,\mathcal M}^{(Q)}
			(\boldsymbol{\kappa})
			=
			\big[
			\overline{\mathbf B}_{\mathcal S,1,K_1}^{(Q)}
			(\boldsymbol{\kappa}_1),\ldots,
			\overline{\mathbf B}_{\mathcal S,\widehat K_d,K_{\widehat K_d}}^{(Q)}
			(\boldsymbol{\kappa}_{\widehat K_d}),
			\overline{\mathbf B}_{\mathcal S,\mathrm{nui}}^{(Q)}
			\big].
		\end{aligned}
		\label{eq:joint_target_interference_matrix}
	\end{equation}
	
	The first $\widehat K_d$ blocks are the candidate co-bin path blocks,
	which contain the principal samples and the fractional Doppler leakage
	retained within the local window for the paths represented by the
	delay classes.
	The final block is the nuisance block. This block collects the retained structured leakage outside the candidate blocks. The retained leakage includes dominant leakage between different pilots and strong leakage from paths outside the candidate blocks. The orthogonal projector onto the column space of the augmented model matrix is
	\begin{equation}
		\overline{\mathbf P}_{\mathcal S,\mathcal M}^{(Q)}
		(\boldsymbol{\kappa})
		=
		\overline{\mathbf A}_{\mathcal S,\mathcal M}^{(Q)}
		(\boldsymbol{\kappa})
		\left[
		\overline{\mathbf A}_{\mathcal S,\mathcal M}^{(Q)}
		(\boldsymbol{\kappa})
		\right]^\dagger .
		\label{eq:augmented_model_projector}
	\end{equation}
	
	The fractional Doppler vector associated with each candidate model is
	estimated by minimizing the concentrated residual energy
	\begin{equation}
		\begin{aligned}
			&\widehat{\boldsymbol{\kappa}}_{\mathcal M}
			={}
			\underset{
				\boldsymbol{\kappa}\in\mathcal K_{\mathcal M}
			}{\arg\min}
			\left\|
			\left[
			\mathbf I_{N_{\mathrm{st}}}
			-
			\overline{\mathbf P}_{\mathcal S,\mathcal M}^{(Q)}
			(\boldsymbol{\kappa})
			\right]
			\overline{\mathbf y}_{\mathcal S}^{(Q)}
			\right\|_2^2,\\
			&\operatorname{RSS}_{\mathcal M}
			={}
			\left\|
			\left[
			\mathbf I_{N_{\mathrm{st}}}
			-
			\overline{\mathbf P}_{\mathcal S,\mathcal M}^{(Q)}
			(\widehat{\boldsymbol{\kappa}}_{\mathcal M})
			\right]
			\overline{\mathbf y}_{\mathcal S}^{(Q)}
			\right\|_2^2 ,
		\end{aligned}
		\label{eq:near_principal_fractional_doppler_estimate}
	\end{equation}
	where $\mathcal K_{\mathcal M}$ denotes the ordered search space for
	$\boldsymbol{\kappa}$. Candidate models that do not satisfy
	$N_{\mathrm{st}}>K_{\mathcal M}+K_{\mathrm{ext}}$ or whose model matrix
	does not have full rank are excluded. The residual
	$\operatorname{RSS}_{\mathcal M}$ obtained in \eqref{eq:near_principal_fractional_doppler_estimate} is used to select the path number using by BIC
	\begin{equation}
		\begin{aligned}
			&\operatorname{BIC}(\mathcal M)
			={}
			N_{\mathrm{st}}\log\!\left(
			\frac{\operatorname{RSS}_{\mathcal M}}{N_{\mathrm{st}}}
			\right)
			+3K_{\mathcal M}\log N_{\mathrm{st}},\\
			&\widehat{\mathcal M}
			={}
			\underset{\mathcal M\in\mathfrak M}{\arg\min}\,
			\operatorname{BIC}(\mathcal M).
		\end{aligned}
		\label{eq:class_bic_order}
	\end{equation}
	
	The penalty counts only the candidate co-bin paths. Each path introduces
	one real fractional Doppler offset and one complex gain, corresponding to
	three real parameters. The nuisance columns are fixed for all candidate
	models and therefore add only a common penalty. If $\widehat{\mathcal M} =(\widehat K_1,\ldots,\widehat K_{\widehat K_d})$, the estimated
	physical path number in the selected co-bin is
	\begin{equation}
		\widehat K_c
		=
		K_{\widehat{\mathcal M}}
		=
		\sum_{g=1}^{\widehat K_d}\widehat K_g,
		\label{eq:total_cobin_path_number}
	\end{equation}
	where $\widehat K_d$ counts the resolved delay class obtained from
	the first stage, whereas $\widehat K_c$ counts the Doppler resolvable
	physical paths represented by these delay classes. The selected fractional Doppler estimate is the NMLL-MLE result corresponding to the model selected by BIC
	\begin{equation}
		\widehat{\boldsymbol{\kappa}}
		=
		\widehat{\boldsymbol{\kappa}}_{\widehat{\mathcal M}} .
		\label{eq:selected_class_fractional_doppler}
	\end{equation}
	
	The path and retained nuisance gains are jointly recovered from the
	converged fractional Doppler estimates as
	\begin{equation}
		\begin{bmatrix}
			\widehat{\mathbf h}\\
			\widehat{\mathbf h}_{\mathrm{nui}}
		\end{bmatrix}
		=
		\left[
		\overline{\mathbf A}_{\mathcal S,\widehat{\mathcal M}}^{(Q)}
		(\widehat{\boldsymbol{\kappa}})
		\right]^\dagger
		\overline{\mathbf y}_{\mathcal S}^{(Q)} .
		\label{eq:joint_gain_estimate}
	\end{equation}
	
	The estimated parameter triplet of the $j$th path represented by the
	$g$th resolved delay class is
	$(\widehat d_g,k_c+\widehat\kappa_{g,j},\widehat h_{g,j})$,
	with $g=1,\ldots,\widehat K_d$ and $j=1,\ldots,\widehat K_g$. 
	The resulting path parameters are used as observations for subsequent target state estimation and as channel parameters for DPF domain channel reconstruction and data recovery.

	\section{CRLB Analysis}
	
	The CRLB is derived as a conditional benchmark for parameter estimation within a co-delay-Doppler bin. It evaluates the estimation accuracy after the hierarchical model order determination has identified the resolved delay representation and the number of physical paths within each resolved delay class. For the
	detected integer Doppler index, the total normalized Doppler of path
	$(g,r)$ is $\nu_{g,r}=k_c+\kappa_{g,r}$. Since $k_c$ is fixed locally, the derivatives with respect to $\nu_{g,r}$ and $\kappa_{g,r}$ are identical. Substituting the uniformly spaced pilot sequence into the DPF domain input output relationship yields
	\begin{equation}
		y_p(m)=\sum_{g=1}^{K_d}\sum_{r=1}^{K_g}h_{g,r}
		\sum_{\upsilon=0}^{\zeta-1}B_{\upsilon,g,r}(m)
		+\widetilde w(m).
		\label{eq:crlb_observation}
	\end{equation}
	
	The expectation of \eqref{eq:crlb_observation} is denoted by $\mu(m;\boldsymbol\theta)=\mathbb E\{y_p(m)\mid\boldsymbol\theta\}$. Since the transformed AWGN has zero mean, the mean received signal is given by its noise free component. Hence, $y_p(m)=\mu(m;\boldsymbol\theta)+\widetilde w(m).$ The real parameter vector is defined as
	$\boldsymbol\theta=[\boldsymbol d^{\mathrm T},\boldsymbol\nu^{\mathrm T},
	\boldsymbol h_{\mathrm R}^{\mathrm T},\boldsymbol h_{\mathrm I}^{\mathrm T}]^{\mathrm T}$,
	where $\boldsymbol d=[d_1,\ldots,d_{K_d}]^{\mathrm T}$, and $\boldsymbol\nu$, $\boldsymbol h_{\mathrm R}$, and $\boldsymbol h_{\mathrm I}$ stack $\nu_{g,r}$, $\Re\{h_{g,r}\}$, and $\Im\{h_{g,r}\}$ over all physical paths. For $\boldsymbol y\sim\mathcal{CN}(\boldsymbol\mu,\sigma_s^2\boldsymbol I_{N_c})$, the logarithm of the likelihood function, excluding constants independent of $\boldsymbol\theta$, is $-\|\boldsymbol y-\boldsymbol\mu\|_2^2/\sigma_s^2$. The $(a,b)$th entry of the Fisher information matrix (FIM) is
	\begin{equation}
		[\boldsymbol J]_{a,b}=\frac{2}{\sigma_s^2}\Re\!\left\{
		\sum_{\xi\in\mathcal S}\sum_{z\in\mathcal Z_Q}
		\left(\frac{\partial\mu(m)}{\partial\theta_a}\right)^{\!*}
		\left(\frac{\partial\mu(m)}{\partial\theta_b}\right)
		\right\}.
		\label{eq:crlb_fim}
	\end{equation}
	
	To evaluate \eqref{eq:crlb_fim}, define the leakage index associated with the $\upsilon$th pilot as
	$q_{\upsilon}(m)=[m-m_{\mathrm{pilot},\upsilon}-k_c]_N$. The corresponding path contribution is
	\begin{equation}
		\begin{aligned}
			B_{\upsilon,g,r}(m)
			&\triangleq \sqrt{E_{\mathrm{pilot},\upsilon}}\,
			\Gamma_N\!\left(q_{\upsilon}(m),\nu_{g,r}-k_c\right)
			e^{\Phi_{\upsilon,g,r}(m)},\\
			\Phi_{\upsilon,g,r}(m)
			&\triangleq \mathrm j\frac{\pi}{N}[k_c+q_{\upsilon}(m)]^2
			-\mathrm j\frac{2\pi m}{N}
			[d_g+k_c+q_{\upsilon}(m)].
		\end{aligned}
		\label{eq:crlb_path_component}
	\end{equation}
	
	Thus, $\mu(m)=\sum_g\sum_r h_{g,r}\sum_{\upsilon}B_{\upsilon,g,r}(m)$. The derivatives with respect to the real and imaginary parts of the complex gain are
	\begin{equation}
		\begin{aligned}
			\frac{\partial\mu(m)}{\partial h_{\mathrm R,g,r}}
			=\sum_{\upsilon=0}^{\zeta-1}B_{\upsilon,g,r}(m),
			\frac{\partial\mu(m)}{\partial h_{\mathrm I,g,r}}
			=\mathrm j\sum_{\upsilon=0}^{\zeta-1}B_{\upsilon,g,r}(m).
		\end{aligned}
		\label{eq:crlb_gain_derivatives}
	\end{equation}
	
	The paths represented by delay class $g$
	share the representative delay parameter $d_g$. Therefore, the derivative
	with respect to $d_g$ collects the contributions of paths assigned to this class:
	\begin{equation}
		\frac{\partial\mu(m)}{\partial d_g}
		=-\mathrm j\frac{2\pi m}{N}
		\sum_{r=1}^{K_g}h_{g,r}
		\sum_{\upsilon=0}^{\zeta-1}B_{\upsilon,g,r}(m).
		\label{eq:crlb_delay_derivative}
	\end{equation}
	
	Conditioned on the detected $k_c$, both $m$ and $q_{\upsilon}(m)$ are fixed discrete indices. Therefore, the Doppler derivative acts only on the Dirichlet leakage coefficient, and no derivative of a periodic impulse is introduced. The derivative with respect to the total normalized Doppler is
	\begin{equation}
		\frac{\partial\mu(m)}{\partial\nu_{g,r}}
		=h_{g,r}\sum_{\upsilon=0}^{\zeta-1}
		\dot B_{\upsilon,g,r}(m),
		\label{eq:crlb_doppler_derivative}
	\end{equation}
	where
	\begin{equation}
		\dot B_{\upsilon,g,r}(m)
		\triangleq \sqrt{E_{\mathrm{pilot},\upsilon}}\,
		\Gamma_N'\!\left(q_{\upsilon}(m),\nu_{g,r}-k_c\right)
		e^{\Phi_{\upsilon,g,r}(m)}.
		\label{eq:crlb_component_derivative}
	\end{equation}
	
	Using $\Gamma_N(q,\kappa)=N^{-1}\sum_{n=0}^{N-1}
	\exp[\mathrm j2\pi n(\kappa-q)/N]$, its derivative is
	\begin{equation}
		\Gamma_N'(q,\kappa)
		=\frac{\mathrm j2\pi}{N^2}\sum_{n=0}^{N-1}n
		\exp\!\left[\mathrm j\frac{2\pi}{N}n(\kappa-q)\right].
		\label{eq:crlb_kernel_derivative}
	\end{equation}
	
	Substituting \eqref{eq:crlb_gain_derivatives}-\eqref{eq:crlb_kernel_derivative} into \eqref{eq:crlb_fim} yields the FIM. The covariance matrix of any unbiased estimator satisfies
	\begin{equation}
		\operatorname{CRLB}(\theta_a)
		=
		\left[
		\boldsymbol J^{-1}(\boldsymbol\theta)
		\right]_{a,a}.
		\label{eq:crlb_final}
	\end{equation}

	From \eqref{eq:crlb_fim}-\eqref{eq:crlb_kernel_derivative},
	the Fisher information depends on the number of pilot blocks
	$\zeta$ and the number of retained leakage samples $N_Q$.
	For a valid coherent observation model, increasing $\zeta$
	generally improves the delay and Doppler bounds by providing more
	pilot phase samples and stacked leakage observations.

	\section{Simulation Results and Analysis}



	In this section, we evaluate the reliability of the proposed TSUR framework, the root mean square errors (RMSEs) of delay and fractional Doppler estimation, the normalized mean square error (NMSE) of channel reconstruction, and the bit error rate (BER) performance under different pilot configurations. Unless otherwise specified, the simulation parameters are set according to Table~\ref{tab:simulation_parameters}.

	\begin{table}[!t]
		\centering
		\caption{Simulation Parameters}
		\label{tab:simulation_parameters}
		\scriptsize
		\renewcommand{\arraystretch}{1.08}
		\setlength{\tabcolsep}{3.5pt}
		\begin{tabular}{p{0.54\columnwidth}p{0.34\columnwidth}}
			\toprule
			\textbf{Parameter} & \textbf{Value} \\
			\midrule
			DPF-domain frame size & $N=1024$ \\
			Reserved pilot index range & $n=0,\ldots,255$ \\
			Number of protected pilot blocks & $\zeta=16$ \\
			First pilot center index & $m_{\mathrm{pilot},0}=8$ \\
			Pilot center interval & $D_{\mathrm{pilot}}=16$ \\
			Selected integer Doppler index & $k_c=1$ \\
			Fractional Doppler vector & $\boldsymbol\kappa=[-0.1,0.2]^{\mathrm T}$ \\
			Continuous normalized delay vector & $\mathbf d=[9.1,9.2]^{\mathrm T}$ \\
			Normalized path-gain profile & $\Delta P=3~\mathrm{dB},\ \lVert\mathbf h\rVert_2^2=1$ \\
			\bottomrule
		\end{tabular}
	\end{table}

	\subsection{Hierarchical Detection Reliability}

\begin{figure}[!t]
		\centering
		\includegraphics[width=3in]{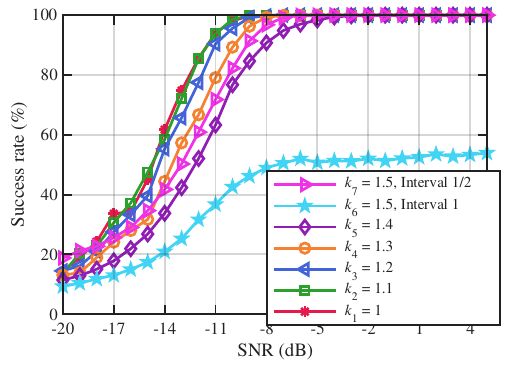}
		\caption{Success rate of integer Doppler localization versus SNR. For $k_7=1.5$, ``Interval 1/2'' means $\hat{k}_c\in\{1,2\}$, while ``Interval 1'' means $\hat{k}_c=1$.}
		\label{fig:IDL_SR}
	\end{figure}

	Before evaluating the co-bin resolution performance, we evaluate the success rate of accurately estimating the integer part of the Doppler. Accurate localization of the integer Doppler index determines the corresponding received pilot locations, which are then used to construct the principal delay observation and the local leakage observation. The tested paths share the same integer Doppler index and have different fractional offsets. The SNR ranges from $-20$ to $5$ dB in $2$ dB increments with 10000 Monte Carlo trials at each point. The experiment is considered successful if the cross-correlation method employed can correctly identify the interval containing the Doppler shift. Figure~\ref{fig:IDL_SR} shows that the closer the Doppler value is to the interval boundary, the greater the fractional Doppler leakage and the lower the success rate of accurately estimating the integer part of the Doppler under the single interval decision criterion. Fractional Doppler values located at interval boundaries can be searched for by examining the two adjacent boundaries separately. Accurate estimation of the integer Doppler index identifies the corresponding pilot locations at the receiver, which are used to construct the delay and local leakage observations.

	\begin{figure}[!t]
	\centering
	\includegraphics[width=3in]{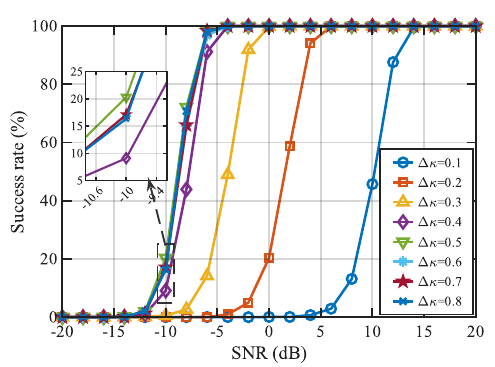}
	\caption{Success rate versus SNR for different fractional Doppler separations within the same delay class.}
	\label{fig:cuss}
	\end{figure}
	
	We examine how the Doppler separation within a co-bin and the leakage caused by fractional Doppler affect the success rate of BIC. The Doppler of the second path is varied within a co-bin, and the probability of correctly identifying the two physical paths is evaluated. Within the same delay class, the two paths have fractional Doppler offset $\kappa_1=-0.4$ and $\kappa_2=-0.4+\Delta\kappa$, where $\Delta\kappa$ varies from $0.1$ to $0.8$. A trial is successful when BIC estimates the true physical path number. Figure~\ref{fig:cuss} shows that the detection probability improves as the SNR increases. As the paths become more distinguishable, the BIC can identify the path model at progressively lower SNRs. On the other hand, the positions of the fractional Doppler offset relative to the bin boundary determine the degree of overlap between the two leakage patterns. A stronger overlap makes the two paths more difficult to distinguish.  The success rate versus SNR for different fractional Doppler separations provides the basis for subsequently estimating the fractional Doppler offsets.

	\subsection{Validation of Delay Class Estimation}

	This experiment isolates the delay class estimation module in the first stage of TSUR. 
	Two well separated delay classes are considered with 
	$\mathbf d=[9.1,\,23.5]^{\mathrm T}$ and 
	$\boldsymbol{\kappa}=[-0.1,\,0.2]^{\mathrm T}$. 
	Since the two delays are far apart, this experiment is not intended to demonstrate co-bin path resolution. 
	Instead, it validates the ability of FBSS-MUSIC and Newton refinement to estimate resolvable delay classes from the principal pilot observations. 
	MUSIC, FBSS-MUSIC, FBSS-MUSIC-Newton, and ZP-PCTD~\cite{Arous2026PCTD} are compared. 
	The SNR ranges from $0$ to $40$ dB in increments of $2$ dB, and each point is averaged over 5000 Monte Carlo trials. 
	FBSS-MUSIC-Newton uses the FBSS-MUSIC estimates to initialize the Newton refinement.
	ZP-PCTD uses zero padding factors of $z_p=8$ and $z_p=16$ to refine the delay grid. 
	It uses the same number of pilots and pilot guard intervals as the proposed method.
	
\begin{figure}[!t]
		\centering
		\includegraphics[width=3in]{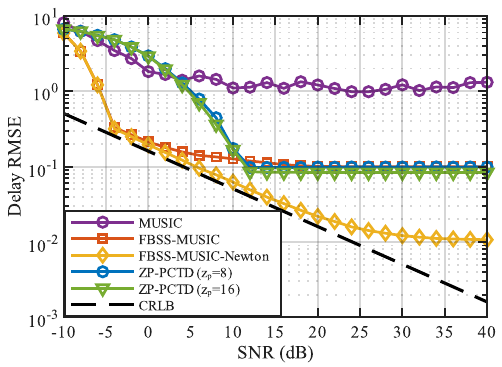}
		\caption{Delay estimation performance for two resolvable delay classes.}
		\label{fig:delay_est}
	\end{figure}

	As shown in Fig.~\ref{fig:delay_est}, the MUSIC exhibits a pronounced error floor due to the rank deficiency of the covariance matrix caused by coherent observations. FBSS restores the effective rank of the signal subspace and substantially improves the delay estimation accuracy. At high SNR, the accuracy of FBSS-MUSIC is limited by the resolution of its delay search grid, which produces a residual error floor. Newton refinement improves the coarse delay estimates beyond the discrete search grid, thereby reducing grid induced quantization errors.  A residual error floor remains at high SNR because of interference from the unknown data symbols. ZP-PCTD exhibits a similar floor because its accuracy is limited by the PCTD delay search discretization and the interference from the unknown data symbols. The latter arises because ZP-PCTD adopts the same block structure as the proposed method, in which the pilot symbols are placed in the leading part of each block and followed by data symbols.

	\subsection{Fractional Doppler Resolution within a Delay Class}
	\begin{figure}[!t]
		\centering
		\includegraphics[width=3in]{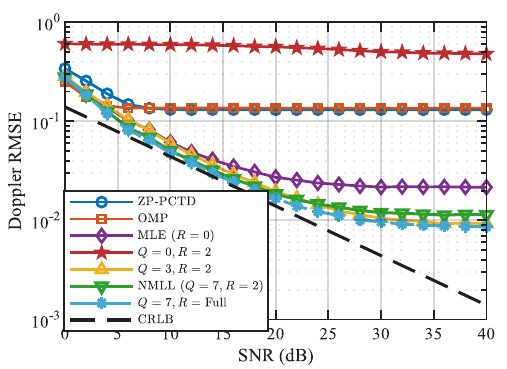}
		\caption{Doppler estimation RMSE of the compared
			methods versus SNR.}\label{fig5}
		\label{fig:dopper_estsamedelay}
	\end{figure}
	
	\begin{figure}[!t]
		\centering
		\includegraphics[width=3in]{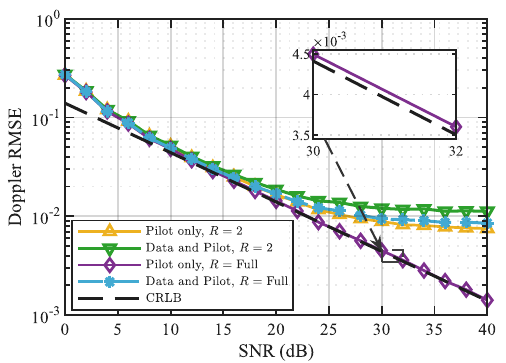}
		\caption{Effect of unknown data symbols on fractional Doppler estimation.}
		\label{fig:pilot_only}
	\end{figure}

	\begin{figure*}[!t]
		\centering
		\includegraphics[width=0.92\textwidth]{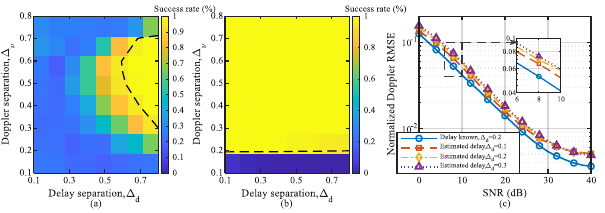}
		\caption{Analysis of fractional Doppler estimation for different delay paths within the same delay class. 
			(a) Success rate of resolving two delay paths using MDL.
			(b) Success rate of detecting two physical paths using BIC.
			(c) Fractional Doppler estimation performance under different delay separations.}
		\label{fig:doppler_delaymiss}
	\end{figure*}

	The fractional Doppler estimation performance is evaluated for two physical paths that share the same delay class. The two paths share the delay $d_1=d_2=9.2$ and have fractional Doppler offsets $\boldsymbol{\kappa}=[-0.4,0.3]^{\mathrm T}$. The SNR ranges from $0$ to $40$ dB in $2$ dB increments, with $3000$ Monte Carlo trials per point. ZP-PCTD and OMP~\cite{Tropp2007OMP} are included as representative sequential
	estimators. OMP uses the same $Q=7$ local observations as the MLE ($R=0$). For NMLL-MLE,
	$Q\in\{0,3,7\}$ at $R=2$ and
	$R\in\{0,2,\mathrm{Full}\}$ at $Q=7$ are considered to evaluate
	the accuracy and complexity tradeoff. As shown in Fig.~\ref{fig:dopper_estsamedelay}, the configuration using only the principal sample with $Q=0$ exhibits a large error floor that is nearly insensitive to SNR. This result shows that the principal samples alone cannot provide accurate fractional Doppler estimates for the considered co-bin paths.  Both OMP and ZP-PCTD saturate at approximately the same RMSE level, which is attributed to the combined effect of path extraction and the leakage of unknown data symbol energy into the pilot observations. The MLE with $R=0$ reduces the estimation error but retains the high SNR floor because the cross pilot contributions are omitted. Modeling the neighboring pilot blocks with $R=2$ further improves the fractional Doppler estimation accuracy. The comparable performance obtained with $Q=3$ and $Q=7$ indicates that most useful fractional Doppler information is concentrated around the principal sample. 
	
	Fig.~\ref{fig:pilot_only}  shows the effect of
	unknown data symbols on fractional Doppler estimation. The results with only pilots are compared
	with those obtained from a frame containing both pilots and unknown
	QPSK data. For the pilot only case, the curve with
	$R=\mathrm{Full}$ closely follows the CRLB at high SNR, while
	$R=2$ keeps a small residual because only the dominant
	interactions among pilots are retained. When unknown data symbols
	are present, both $R=2$ and $R=\mathrm{Full}$ exhibit a nonzero
	high SNR floor. It shows that the residual floor in the
	data bearing frame is mainly caused by data symbol leakage into the
	selected pilot windows, rather than by the fractional Doppler search
	itself. The parameters $Q$ and $R_{\mathrm p}$ influence the complexity of
	the fractional Doppler estimation. The overall computational complexity is approximated by $\mathcal O\!\left(
		I_{\kappa,\mathbf r}\left[K_{\mathbf r}N_QS_{R_{\mathrm p}}+N_{\mathrm{st}}M_{\mathbf r}^{2}+M_{\mathbf r}^{3}\right]\right)$, where $I_{\kappa,\mathbf r}$ denotes the number of iterations.  Moreover,
		$N_{\mathrm{st}}=\zeta N_Q$ is the stacked observation
		length, $K_{\mathbf r}$ is the candidate number of target paths,
		and $M_{\mathbf r}=K_{\mathbf r}+K_{\mathrm{ext}}$ is the total
		number of target and retained nuisance columns. The quantity $S_{R_{\mathrm p}}$ denotes the number of transmitted
		pilot contributions retained across all $\zeta$ received pilot windows. The first term represents the construction of the target leakage
		matrix and is directly controlled by both $Q$ and $R_{\mathrm p}$.
		The second term accounts for the matrix multiplications required in the concentrated least squares fit, while the third term corresponds to solving the resulting $M_{\mathbf r}\times M_{\mathbf r}$ linear
		system. For the two path case with $\zeta=16$ and $Q=7$, setting
		$R_{\mathrm p}=2$ reduces the overall computational cost by
		approximately $63\%$ compared with full modeling of the interactions among pilots.
		Together with the small performance gap in Fig.~\ref{fig:dopper_estsamedelay}, this confirms
		that the dominant cross pilot interactions can be captured with
		substantially lower complexity.

	The fractional Doppler estimation performance is evaluated for physical paths with different continuous delays within the same delay class. After integer Doppler index and delay refinement, NMLL-MLE in the second stage recovers the fractional Doppler offsets from the stacked local leakage samples. For this experiment, the first path is fixed at
	$(d_1,\nu_1)=(8.6,0.6)$. The delay and Doppler coordinates of the second path are independently varied as
	$d_2\in\{8.7,8.8,\ldots,9.4\}$ and
	$\nu_2\in\{0.7,0.8,\ldots,1.4\}$, respectively. Accordingly, the delay separation $\Delta_d=d_2-d_1$ and the Doppler separation $\Delta_\nu=\nu_2-\nu_1$ both range from $0.1$ to $0.8$ with an interval of $0.1$.
	TSUR is evaluated at $\mathrm{SNR}=10$ dB using $3000$ Monte
	Carlo trials at each grid point. The black dashed contours mark
	the $95\%$ success probability boundaries for MDL and BIC,
	respectively. Figure~\ref{fig:doppler_delaymiss} evaluates the robustness of TSUR when two closely different delays are represented by one delay class. Fig.~\ref{fig:doppler_delaymiss}(a)
	shows that the MDL result depends on both the delay separation and
	the diversity of the leakage responses. Closely spaced paths may produce highly correlated covariance signatures across the pilot blocks when their Doppler offsets lie near the two boundaries of the same Doppler interval. Fig.~\ref{fig:doppler_delaymiss}(b) shows that the BIC path number decision is governed mainly by the fractional Doppler separation after the paths are represented by one resolved delay class. Fig.~\ref{fig:doppler_delaymiss}(c) shows the fractional Doppler estimation accuracy after the two paths have been resolved by BIC. The Doppler coordinates are fixed as $\boldsymbol{\nu}=[0.9,1.2]$, and the case with known true path delays is used as the oracle benchmark. The fractional Doppler estimation accuracy improves as the SNR increases for all curves. The oracle benchmark provides the best performance because the true path delays are known in advance. When the delay difference between the two paths ranges from $0.1$ to $0.3$, only one delay class is resolved in the first stage. Under this delay mismatch, the fractional Doppler estimation performance degrades by approximately $4$ dB compared with the oracle benchmark. 

	\begin{figure}[!t]
		\centering
		\includegraphics[width=3in]{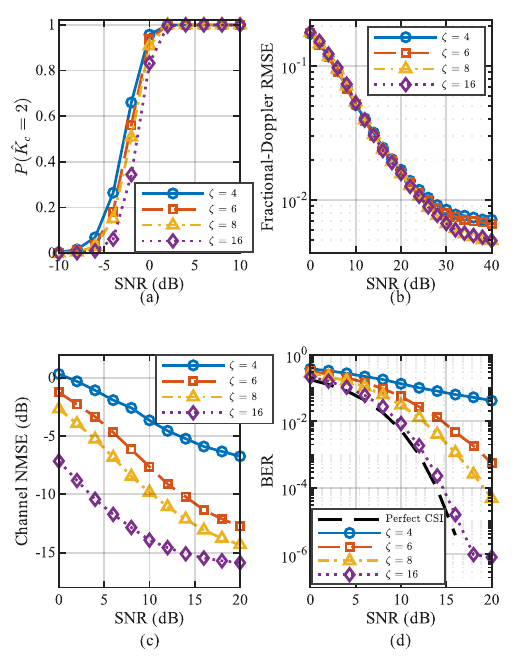}
		\caption{Effect of pilot count on sensing and communication performance under fixed total pilot energy. 
			(a) BIC detection of the number of paths. 
			(b) Fractional Doppler estimation performance. 
			(c) Channel reconstruction performance. 
			(d) BER of QPSK detection. }
		\label{fig7}
	\end{figure}
	\subsection{Pilot Effect on Sensing and Communication Performance}

	Fig.~\ref{fig7} evaluates the impact of the number of uniformly spaced pilot blocks on sensing and communication performance under a fixed total pilot energy. 
	Changing $\zeta$ affects both the number of phase progression and local leakage observations and the energy assigned to each pilot. 
	The sensing receiver observes a frame containing both pilots and unknown QPSK data, while FBSS-MUSIC, BIC, and NMLL-MLE use only the known pilot model.
	 Fig.~\ref{fig7}(a) shows that using fewer pilot blocks gives a larger
	pilot amplitude and can slightly advance the BIC transition at low
	SNR. However, this advantage does not translate into better parameter
	or channel estimation. As shown in Fig.~\ref{fig7}(b), increasing $\zeta$
	reduces the high SNR fractional Doppler offsets RMSE floor because more pilot
	blocks provide a larger aperture and more stacked local leakage samples
	for NMLL-MLE. Fig.~\ref{fig7}(c) shows that increasing the number of pilots $\zeta$ improves the channel estimation accuracy. The result indicates that channel reconstruction benefits not only from correct detection of the number of paths but also from more accurate Doppler and gain fitting over a richer pilot observation. Fig.~\ref{fig7}(d) shows that the improved reconstructed CSI directly benefits QPSK data detection. Increasing $\zeta$ moves the BER curve closer to the perfect CSI reference. The pilot count does not bring a uniformly monotonic gain for all sensing tasks under fixed total pilot energy. A small number of high power pilots is more suitable for low SNR path number detection, while a larger number of pilots improves continuous parameter estimation, channel reconstruction, and data detection.

	\section{Conclusion}
	
	This paper investigated co-delay-Doppler bin multipath resolution for communication assisted sensing with DFT-P-OCDM. A DPF domain relationship between the input and output was derived. Based on this relationship, the TSUR framework was developed to estimate the resolved delay classes, the number of physical paths in each class, their fractional Doppler offsets, and complex gains. CRLBs were derived to benchmark the attainable estimation accuracy, and the pilot configuration was analyzed to characterize the tradeoff between sensing resolution and communication recovery. Simulation results demonstrated the superiority and robustness of TSUR for co-bin path resolution when the paths have identical delays or small delay differences. These findings indicate that TSUR can serve as an effective front end estimator of path parameters for communication assisted sensing with DFT-P-OCDM.Since the leakage models are waveform specific, extending the co-bin analysis to AFDM requires deriving the corresponding relationship between the input and output, characterizing its leakage behavior, and developing a suitable resolution framework. These issues will be investigated in future work.

\end{document}